\documentclass[hidelinks,a4paper,10pt,onecolumn, switch]{article}
\pdfoutput=1

\usepackage{fancyhdr}
\fancypagestyle{plain}{%
\fancyhf{}
\fancyfoot[R]{\thepage}
\fancyfoot[L]{\small \textit{Preprint}
}}

\usepackage[affil-it]{authblk}
\makeatletter
\def\@maketitle{%
\newpage
\null
\vskip 2em%
\begin{center}%
	\let \footnote \thanks
	{\Large\bfseries \@title \par}%
	\vskip 1.5em%
	{\normalsize
		\lineskip .5em%
		\begin{tabular}[t]{c}%
			\@author
		\end{tabular}\par}%
	\vskip 1em%
	{\normalsize \@date}%
\end{center}%
\par
\vskip 1.5em}
\makeatother

\usepackage[a4paper, total={17.5cm,24cm}]{geometry}

\usepackage[font=normal,labelfont=bf]{caption}

\usepackage{sectsty} 

\usepackage{titlesec}
\titlespacing\section{0pt}{12pt plus 3pt minus 3pt}{1pt plus 1pt minus 1pt}
\titlespacing\subsection{0pt}{10pt plus 3pt minus 3pt}{1pt plus 1pt minus 1pt}
\titlespacing\subsubsection{0pt}{8pt plus 3pt minus 3pt}{1pt plus 1pt minus 1pt}

\titleformat{\section}{\normalfont\large\bfseries}{\thesection}{1em}{}

\titleformat{\subsection}{\normalfont\normalsize\bfseries}{\thesubsection}{1em}{}

\titleformat{\subsubsection}{\normalfont\normalsize}{\thesubsubsection}{1em}{}

\titleformat{\paragraph}[runin]{\normalfont\normalsize\itshape}{\theparagraph}{1em}{}

\usepackage{mathtools,upgreek,amsmath,amssymb}

\usepackage[utf8]{inputenc}	
\usepackage[T1]{fontenc}	
\usepackage{xcolor}		
\usepackage[colorlinks = false]{hyperref}	
\usepackage{booktabs} 		
\usepackage{nicefrac}		
\usepackage{microtype}		
\usepackage{lineno}		
\usepackage{float}			

\usepackage{siunitx}[=v2]
\usepackage{nicefrac,multirow}

\usepackage{subcaption}
\usepackage{algorithmic,algorithm2e}

\usepackage{enumitem}
\setlist[enumerate]{label*=\arabic*.} 

\newcommand\numberthis{\addtocounter{equation}{1}\tag{\theequation}}

\DeclareMathOperator{\tr}{tr}

\newcommand{\inte}[3]{\int \limits_{ #1} #2 \; \mathrm{d} #3}

\newcommand{\nabr}{\nabla_{\ve{X}}}

\newcommand{\diffp}[2]{\frac{\partial #1}{\partial #2}}
\newcommand{\ddiffp}[2]{\dfrac{\partial #1}{\partial #2}}

\newcommand{\ve}[1]{\boldsymbol{#1}} 
\newcommand{\te}[1]{\mathbf #1}
\newcommand{\tg}[1]{\boldsymbol{#1}}

\newcommand{\rset}{\mathbb{R}}

\newcommand{\nv}{\mathrm}
\newcommand{\comma}{\hspace{3mm} \text{,}}
\newcommand{\commaf}{\hspace{3mm} \text{,}}

\newcommand{\glmand}{\hspace{3mm} \text{and} \hspace{3mm}}
\newcommand{\glmwith}{\hspace{3mm} \text{with} \hspace{3mm}}
\newcommand{\commam}{\hspace{3mm} \text{,} \hspace{3mm}}
\newcommand{\point}{\hspace{3mm} \text{.}}

\newcommand{\psii}{\overline \psi}
\newcommand{\psiinn}{\psii^\nv{PANN}}
\newcommand{\psiinno}{\psii^\nv{PANN}_0}
\newcommand{\psiinon}{\psii^\nv{NN}}

\newcommand{\fiso}{\overline{\te F}}
\newcommand{\ciso}{\overline{\te C}}
\newcommand{\biso}{\overline{\te b}}
\newcommand{\omref}{\varOmega_{0}}
\newcommand{\domref}{\partial \varOmega_{0}}
\newcommand{\domrefu}{\partial \varOmega_{0,\hat{\ve u}}}
\newcommand{\domreft}{\partial \varOmega_{0,\hat{\ve t}}}

\newcommand{\veZero}{\textbf{\textit{0}}}

\newcommand{\pki}{\overline{\te P}}

\newcommand{\rs}{\mathbb{R}}
\newcommand{\rsnn}{\rs_{\geq 0}}
\newcommand{\rsp}{\rs_{> 0}}
\newcommand{\so}{\mathbb{SO}(3)}

\newcommand{\tes}{\mathcal{T}^2}
\newcommand{\tesp}{\mathcal{T}^2_+}
\newcommand{\tesu}{\mathcal{T}_\nv{u}^2}
\newcommand{\tess}{\mathcal{T}_\nv{s}^2}

\newcommand{\pp}{\tilde p}

\usepackage{multicol}

\DeclareMathOperator{\diag}{diag}
\DeclareMathOperator{\cof}{cof}
\DeclareMathOperator{\mse}{MSE}
\DeclareMathOperator*{\stat}{stat}
\DeclareMathOperator*{\argmin}{arg min}

\newcommand{\is}{(\ici,\iici)}
\newcommand{\isc}{\ve{\tilde{{\mathcal{I}}}}}

\newcommand{\ici}{\overline{I}_1}
\newcommand{\iici}{\overline{I}_2}
\newcommand{\iiici}{\overline{I}_3}
\newcommand{\ipci}{\tilde{I}_1}
\newcommand{\ipcii}{\tilde{I}_2}

\newcommand{\afu}{\mathcal{A}}

\DeclareEmphSequence{\bfseries}

\usepackage{mdframed}

\usepackage{amsthm}

\theoremstyle{definition}
\newtheorem{definition}{Definition}
\newtheorem{theorem}{Theorem}
\newtheorem{lemma}[theorem]{Lemma}

\newtheorem{corollary}[theorem]{Corollary}

\theoremstyle{remark}
\newtheorem{remark}{Remark}

\title{When invariants matter: The role of I1 and I2 in neural network models of incompressible hyperelasticity
}

\begin{document}

\author[a]{Franz Dammaß}
\author[a]{Karl A. Kalina}
\author[,a,b]{Markus Kästner \thanks{Contact: \texttt{markus.kaestner@tu-dresden.de} }}
\affil[a]{Institute of Solid Mechanics, TU Dresden, Germany}
\affil[b]{DCMS -- Dresden Center for Computational Materials Science, Dresden, Germany}
\date{}
\maketitle

\begin{abstract}
	Machine learning-based approaches enable flexible and precise material models.
	For the formulation of these models, the usage of invariants of deformation tensors is attractive, since this can a priori guarantee objectivity and material symmetry.
	In this work, we consider incompressible, isotropic hyperelasticity, where two invariants $\ici$ and $\iici$ are required for depicting a deformation state.
	First, we aim at enhancing the understanding of deformation invariants. We provide an explicit representation of the set of invariants that are actually admissible, i.e. for which $(\ici,\iici) \in \rset^2$ a physical deformation state does indeed exist. Furthermore, we prove that uniaxial and equi-biaxial deformation states correspond to the boundary of the set of admissible invariants.
	Second, we study how the experimentally-observed constitutive behaviour of different
	materials can be captured by means of neural network models of incompressible hyperelasticity, depending on whether both $\ici$ and $\iici$ or solely one of the invariants, i.e. either only $\ici$ or only $\iici$, are taken into account. To this end, we investigate three different experimental data sets from the literature.
	In particular, we demonstrate that considering only one invariant --~either $\ici$ or $\iici$~-- can allow for good agreement with experiments in case of small deformations. In contrast, it is necessary to consider both invariants for precise models at large strains, for instance when rubbery polymers are deformed.
	Moreover, we show that multiaxial experiments are strictly required for the parameterisation of models considering $\iici$. Otherwise, if only data from uniaxial deformation is available, significantly overly stiff responses could be predicted for general deformation states.
	On the contrary, $\ici$-only models can make qualitatively correct predictions for multiaxial loadings even if parameterised only from uniaxial data, whereas $\iici$-only models are completely incapable in even qualitatively capturing experimental stress data at large deformations.

	\vspace{5mm}
	\noindent
	\textbf{Keywords: }Incompressible hyperelasticity~\textendash~Deformation invariants~\textendash~Second invariant~\textendash~Physics-augmented neural networks~\textendash~Constitutive model~\textendash~Finite deformations
	
\end{abstract}

\section{Introduction}
\label{sec:intro}

The formulation of material models in terms of deformation invariants is appealing, since it enables to fulfil objectivity, material symmetry and compatibility with the balance of angular momentum in straightforward manner by construction, cf.~\cite[Ch.~8]{silhavy1997}.
For this, the principal invariants of deformation tensors are frequently used. Alternatively, principal stretch-based representations may also be considered, see for instance \cite[Sect.~50.3]{gurtin2010}.
In this work, we focus on incompressible isotropic hyperelasticity, which is relevant for important classes of materials, e.g. rubbery polymers. In this case, two invariants, typically given the symbols $I_1$ and $I_2$ or --~if a volumetric-isochoric split of the deformation gradient is employed~\cite{flory1961}~-- \, $\ici$ and $\iici$, are sufficient for formulating a scalar-valued isotropic tensor function such as the free energy functional.
Apart from this, the concept of models formulated in invariants can also be employed for compressible materials, and extended to anisotropy~\cite{kalina2025}, inelasticity~\cite{rosenkranz2024b} and coupled problems~\cite{gebhart2022,kalina2024}.

In recent years, data-driven~\cite{fuhg2024d,kirchdoerfer2016,wiesheier2024} and machine learning-based approaches have become popular in constitutive modelling. One important advantage of these approaches is a large flexibility when applied to varied complex materials, in particular regarding the capability of capturing highly nonlinear responses as they can be observed in elastomers or biomaterials, for instance.
Among these methods, {neural networks (NNs)} that incorporate fundamental physical principles are very commonly used~\cite{liu2021,dornheim2023,fuhg2024d}.
To this end, \textit{physics-informed~(PINN)}~\cite{raissi2019,henkes2022,bastek2023,harandi2024}, \textit{mechanics-informed} \cite{asad2022}, \textit{physics-augmented~(PANN)}~\cite{klein2022,linden2023,kalina2024,dammass2025}, \textit{physics-constrained}~\cite{kalina2023}, \textit{physics-based}~\cite{aldakheel2025,baktheer2025}, or \textit{constitutive~(CANN)}~\cite{linka2021,linka2023a,holthusen2024} artificial NNs have been established.
In this context, invariant-based NN-ansatzes for hyperelastic potentials have frequently been employed for both compressible~\cite{klein2022,thakolkaran2022,linden2023} and incompressible~\cite{linka2021,linka2023,bahmani2024,kuhl2024,dammass2025,klein2025} elastic materials.
Almost all invariant-based NN models of incompressible hyperelasticity consider two deformation invariants. Nevertheless, a model based only on $I_2$ has recently been introduced~\cite{kuhl2024}.
As alternative, NN potentials directly depending on the coordinates of strain or deformation tensors have also been proposed \cite{asad2022,vlassis2020,vlassis2022}. While this can offer more flexibility, material symmetry may be more complicated to enforce, cf.~\cite{fernandez2020,klein2022,garanger2024}.
More recently, principal stretch-based NN approaches have also been established~\cite{vijayakumaran2025,abdolazizi2025}, which are naturally limited to the case of isotropy.

Apart from NN-based approaches, many classical hyperelastic models are also formulated in invariants, for instance the \textit{Neo-Hookean}, the \textit{Mooney}~\cite{mooney1940} or the \textit{Yeoh}~\cite{yeoh1993} model, see the reviews~\cite{steinmann2012,ricker2023}.
While the most performant models include two invariants, less sophisticated approaches only based on the first invariant ($I_1$ or $\ici$) are also widely used~\cite{ricker2023}, with the Neo-Hookean model being the most prominent one.

Although intensive research has addressed the formulation of hyperelastic models, less efforts have been devoted to the analysis of the deformation invariants as kinematical quantities and their importance for the parameterisation and the design of material models, and in particular of NN-based approaches.
Nevertheless, such understanding may be of great importance for the design of experiments, where it can be crucial to know which values of the invariants are actually physically admissible, i.e. which values correspond to deformation states that can occur in a deformed material.
Furthermore, this is also of relevance for the development of invariant-based models and the discussion of model properties, see e.g.~\cite{dammass2025}.
For instance, if a material model is desired to be predictive for arbitrary load states, it should be in agreement with experimental data for a selection of load states that is representative for the set of all admissible deformation states to a good degree of approximation.
In the literature, there are comparatively few contributions discussing the admissibility of deformation invariants.
For the case of compressible elasticity, admissibility of the deformation invariants has been studied early in 1966 in~\cite{roesel1966} and \cite{agarwal2005}, providing implicit expressions for the admissible set and a graphical representation based on spherical coordinates.
Moreover, necessary conditions for admissibility were defined in~\cite{moriarty1971} and have later been used for polyconvexity proofs in~\cite{hartmann2003b,schroeder2003}, for instance.
For the incompressible case, similar discussions are found in~\cite{sawyers1977} and later~\cite{baaser2013}, with visualisations in the $I_1$-$I_2$ plane.
However, while it is often stated that \textit{Uniaxial tension~(UT)} and \textit{Equi-biaxial tension~(BT)} correspond to limiting load cases, see e.g.~\cite{hart-smith1966,criscione2000,baaser2013} and numerical evidence has been provided for this in the 1960s~\cite{hart-smith1966}, to the best of the authors' knowledge, a proof is unavailable.

The subject of this work is twofold: 
First, we aim at enhancing the understanding of deformation invariants in isochoric hyperelasticity. In particular, we present an explicit representation of the set of invariant tuples $(I_1,I_2)$ that are admissible, i.e. correspond to physically meaningful deformation states. Moreover, we formally prove that the simple scenarios of uniaxial deformation and equi-biaxial deformation correspond to limiting cases, which bound the set of admissible $(I_1,I_2)$.
Second, we study how the experimentally-observed constitutive behaviour of different materials can be captured by means of NN-based models, depending on whether only one of the invariants or both are taken into account.

This paper proceeds as follows. In Sect.~\ref{sec:hyper_fund}, the fundamentals of incompressible hyperelasticity are summarised. Thereafter, in Sect.~\ref{sec:adm-inv}, a formal definition of admissibility of the deformation invariants is given and an explicit representation of the set of admissible invariants is derived. Moreover, simple loading scenarios are discussed and it is proven that these correspond to limiting cases in terms of the invariants.
Then, in Sect.~\ref{sec:fits}, it is studied when and why both deformation invariants should be considered for NN-based modelling of incompressible hyperelasticity.
The paper closes with concluding remarks and an outlook regarding future work in Sect.~\ref{sec:concl}.
Additional mathematical relations as well as data are provided in the Appendix.

 \paragraph{\emph{Notation.}}
 Within this paper, italic symbols are used for scalar quantities ($J$, $\psi$) and bold italic symbols for vectors ($\ve u \in \mathcal T^1 = \rs^3  $).
 For second-order tensors, bold non-italic letters ($\te P, \, \tg \upsigma \in \tes = \rs^3 \otimes \rs^3$) are used. Transpose and inverse of $\te t \in \tes$ are given by $\te t^\top$ and $\te t^{-1}$, respectively. Additionally, $\tr \te t$, $\det \te t$, $\cof \te t = \det(\te t) \, \te t^{-\top}$ are used to indicate trace, determinant and cofactor, respectively.
 A second-order tensor $\te d$ that is represented by a diagonal matrix with $a_1$, $a_2$, $a_3$ on its main diagonal is written as $\te d = \diag(a_1,a_2,a_3)$.
 Furthermore, the sets of $N$-th order tensors are denoted as $\mathcal T^N, \, N \in \mathbb N_{>0}$, and $\tess = \{ \left. \te t \in \tes \right| \te t = \te t^\top\}$, $\tesp = \{ \left. \te t \in \tes \right| \det \te t > 0\}$, $\tesu = \{ \left. \te t \in \tes \right| \det \te t = 1\}$.
 Moreover, the special orthogonal group is given by $(\so, \cdot)$ with $\so = \{ \left. \te Q \in \tes \right| \te Q^{-1} = \te Q^\top \wedge \det \te Q = 1 \}$, and the inner product  $\cdot : (\te C, \te D) \mapsto \te C \cdot \te D = C_{kl} D_{li} \, \ve e_k \otimes \ve e_i$.%
 \footnote{It seems worth mentioning that many concepts and results of matrix theory can be employed in straightforward manner to second order tensors, since they are independent of the metric, and, moreover, Cartesian coordinates are considered here, cf.~e.g.~\cite{hartmann2003b,ciarlet1988}.}
 Therein, $\ve e_k\in \rs^3$ and $\otimes$ denote the $k$-th Cartesian basis vectors and the dyadic product, with the Einstein summation convention applied. The second order identity tensor is denoted as~$\te I$.
 The space of square integrable functions is denoted as $\mathbb L^2$, and $\mathbb H^1$ is the first-order Sobolev space of $\mathbb L^2$ functions of which the derivatives are also in $\mathbb L^2$.
 Finally, the derivative of a quantity $q$ w.r.t. time is written as $\dot q$, and quantities related to the isochoric portion of deformation are marked by an overbar, i.e. $\bar \psi$.
 
 For reasons of readability, the dependency of field quantities on the position in space and on time is usually not written explicitly within this work, for instance, $\pp$ is written instead of $\pp(\ve X, t)$.

\section{Fundamentals of hyperelasticity}
\label{sec:hyper_fund}

\subsection{Kinematics}
Let $\omref \subset \rs^3$ the reference configuration of a solid, and $\varOmega \subset \rs^3$ its deformed configuration. Furthermore, let $T \subset \rset$ the time interval of interest, and $\ve X \in \omref$ the coordinate of a material point in the reference configuration.
The motion is then given by 
\begin{equation}
	\ve \chi : \omref \times T \longrightarrow \varOmega \commaf \hspace{3mm} (\ve X, t) \longmapsto \ve x(\ve X,t)
	\label{eq:chi}
\end{equation} 
which can be assumed to be a homeomorphism, i.e. $\ve \chi$ is bijective and both $\ve \chi$ and $\ve \chi^{-1}$ are continuous.
The displacement of a material point can be obtained as	$\ve u = \ve x - \ve X$,
and the deformation gradient $\te F \in \tesp$ and its determinant $J \in \rsp$ are given by
\begin{equation}
	\te F = \left(\nabr \ve \chi\right)^\top \qquad \text{and} \qquad J= \det \, \te F \commaf
\end{equation}
with $\nabr$ denoting the Nabla operator with respect to the reference coordinate $\ve X$.
Following Flory~\cite{flory1961}, with $\te I \in \tes$ denoting the identity, $\te F$ can be multiplicatively decomposed into 
\begin{equation}
	\te F = \te F^\nv{vol} \cdot \fiso
	\quad \glmwith \quad \te F^\nv{vol} = J^{1/3} \, \te I
	\quad \glmand \quad \fiso = J^{-1/3}\, \te F
	\commaf
	\label{eq:flory}
\end{equation}
such that $\fiso\in \tesu$, i.e. $\det \fiso =1$.
In order to quantify the deformation of the material, we introduce the symmetric positive definite right and left Cauchy-Green deformation tensors $\te C = \te F^\top \cdot \te F \in \tess$ and $\te b = \te F \cdot \te F^\top  \in \tess$.
Similarly, we define their isochoric analogues by
\begin{equation}
	\ciso  = \fiso^\top \cdot \fiso = J^{-2/3} \, \te C 
	\quad \glmand \quad
	\biso = \fiso \cdot \fiso^\top = J^{-2/3} \, \te b \point
	\label{eq:cbar}
\end{equation}
For the formulation of constitutive relations, it can beneficial to use invariants of tensorial kinematic quantities. In this manner, isotropy of material models can be ensured, for instance.
An extension to anisotropic materials can also be realised, defining \textit{additional invariants} depending on the deformation tensor and structural tensors, see e.g.~\cite{holzapfel2000b,kalina2025} and \cite[Sect.~9.3.2]{haupt2002}.
In the following, we limit ourselves to isotropic incompressible hyperelasticity and consider the first and the second principle invariants of $\ciso$ and $\biso$,
\begin{equation}
  \ici = \tr \ciso =  \tr \biso
	\quad \glmand \quad 
	\iici = \tr \cof \ciso =  \tr \cof \biso\commaf
	\label{eq:prinInv}
\end{equation}
which, together with $\iiici = J^2 =1$, are the coefficients of the characteristic polynomial of $\ciso$ and $\biso$,
\begin{equation}
	\kappa^3 - \ici \, \kappa^2 + \iici \, \kappa -1 = 0
	\point
	\label{eq:charpol}
\end{equation}
In the following, we denote the corresponding solutions for $\kappa$, which are the eigenvalues of $\ciso$ and $\biso$ by $\kappa_m$, $m \in \{1,2,3\}$. Due to symmetry, these eigenvalues are real, and positive as shown in Lemma~\ref{lem:posEV} in App.~\ref{sec:posEV}, i.e. $\kappa_m \in \rsp \, \, \forall \,m $.
In terms of $\kappa_m$, the invariants can be represented as
\begin{equation}
	\ici = \kappa_1 + \kappa_2 + \kappa_3 
	\quad \glmand \quad
	\iici = \kappa_1 \, \kappa_2 + \kappa_1 \, \kappa_3 + \kappa_2 \, \kappa_3
	\quad \glmand \quad
	\iiici =  \kappa_1 \, \kappa_2 \, \kappa_3 = 1
	\comma
\end{equation}
and the principal stretches can be computed as $\lambda_m = \sqrt{\kappa_m}$. 
\begin{remark}
In the absence of volumetric deformation, i.e. given a deformation state for which $J = 1$, the isochoric invariants $\ici$, $\iici$ coincide with the principal invariants $I_1 = \tr \te C = \tr \te b$ and $I_2 = \tr \cof \te C = \tr \cof \te b$, i.e. $\ici \equiv I_1$ and $\iici \equiv I_2$.
As a consequence, the results regarding admissible values of the invariants in Sect.~\ref{sec:adm-inv} apply for both $\ici$, $\iici$ and $I_1$, $I_2$ in case of perfect incompressibility.
\end{remark}
\begin{remark}
Although $\ici \equiv I_1$ and $\iici \equiv I_2$ for $J=1$, the decision about which invariants are considered for model formulation can have important implications, for instance regarding the decoupling of isochoric deformations and hydrostatic pressure~\cite{dammass2025}. 
Moreover, when pre-defining $\te F \in \tesu$, the constraint $\det \te F = 1$ may have to be taken into account for differentiation, cf.~\cite[Sect.~3.4]{gurtin2010} and \cite[Sect.~1.2]{ciarlet1988}.
As an alternative, the volumetric-isochoric split can be considered for model formulation as below in Sect.~\ref{sec:psiDef}. Therefore, $\ici$ and $\iici$ are preferred over $I_1$, $I_2$ in this contribution.
\end{remark}

\subsection{Free energy and stress}
\label{sec:psiDef}
In order to describe the incompressible hyperelastic material response, we introduce the field variable 
\begin{equation}
	\pp : \omref \times T\longrightarrow \rs
	\quad,\quad
	\left(\ve X, t\right) \longmapsto \pp\left(\ve X, t\right)
	\label{eq:pfield} 
\end{equation} 
of pressure type, which is assumed to form the set of independent state variables together with the deformation, cf.~\cite{chang1991,holzapfel2000,wriggers2008}.
The free energy functional is then defined as
\begin{equation}
	\psi : \tesp \, \times \, \rs 
	\longrightarrow \rsnn \quad,\quad 
	(\te F, \pp) \longmapsto \psi(\te F, \pp) \comma
	\label{def:psi-gen}
\end{equation}
and an additive decomposition
\begin{equation}
	\psi(\te F, \pp) = \psii (\fiso) + \pp \, (J-1)  \point
	\label{def:psi-split}
\end{equation}
is adopted based on the postulate of Flory~\cite{flory1961}, in line with our previous work~\cite{dammass2025}. Therein, the second term accounts for incompressibility, and $\pp$ can be interpreted as a Lagrangian multiplier, whereas the free energy due to isochoric deformation is expressed by
	$\psii : \tesu \rightarrow \rsnn,\, \fiso \mapsto \psii (\fiso)$.
From the free energy functional $\psi$, the first Piola-Kirchhoff stress $ \te P \in \tes$ follows as
\begin{equation}
	\te P = \diffp{\psi}{\te F} 
	= \diffp{\psii(\fiso)}{\te F} + \pp \, J \, \te F^{-\top}
	\label{eq:PK1}
	\point
\end{equation} 
Moreover, the Cauchy stress $\tg \upsigma \in \tess$ is obtained by means of a \textit{push forward} to the deformed configuration
\begin{equation}
 \tg \upsigma = \frac{1}{J} \, \te P  \cdot \te F^{\top} \comma
 \label{eq:pushForw}
\end{equation}
from which the hydrostatic pressure can be identified as \mbox{$p = - {1}/{3} \tr \tg \upsigma = - {1}/({3 \, J}) \, \te P : \te F^\top$}, see e.g.~\cite[p.~125]{holzapfel2000}. For the specific model considered in this work, the hydrostatic pressure can be rewritten as $p = - \pp$, cf.~\cite[App.~A.1]{dammass2025}.%
\footnote{Note that in general $p \neq - \pp$ for models not incorporating the volumetric-isochoric split of $\te F$. Instead, contributions to hydrostatic pressure can also arise from purely isochoric deformations, see~\cite[App.~A]{dammass2025}.}
	
\subsection{Balance laws, incompressibility and thermodynamic consistency}
Governing equations of incompressible hyperelasticity can be derived demanding stationarity of the energy stored in the solid, for which the energy potential
\begin{align}
	&\varPi: M \times \mathbb L^2(\omref) \longrightarrow \mathbb R
	\quad , \quad
	(\ve \chi, \pp) \longmapsto \varPi(\ve \chi, \pp) = \inte{\omref}{ \psi }{V} - \varPi^\nv{ext}
	\comma
\end{align}
can be defined. Let the potential of external forces
\begin{equation}
	\varPi^\nv{ext} = - \inte{\domreft}{\hat{\ve{t}} \cdot {\ve \chi} }{A}
	\commaf
\end{equation}
with $\hat{\ve{t}}$ denoting the Piola traction vector prescribed on $\domreft \subset \domref$ with $\domreft \cap \domrefu = \varnothing$. The set of admissible motions is given by
\begin{equation}
	M = \left\{ \ve \chi \in \left.\left[\mathbb H^1(\omref)\right]^3 \right\arrowvert \ve \chi(\ve X) - \ve X = \hat{\ve{u}}_t \, \forall \ve X \in \domrefu  \right\} \comma
\end{equation}
wherein $\hat{\ve{u}}_t$ denote the displacements prescribed on the respective Dirichlet part of the boundary, $\domrefu \subset \domref$.
Demanding stationarity
\begin{equation}
	\varPi \longrightarrow \stat\limits_{\ve \chi, \pp}
	\commaf
	\label{eq:varPrin} 
\end{equation}
and employing standard arguments of calculus of variations, we obtain the balance of linear momentum
\begin{equation}
	\nabr \cdot \te{P}^\top = \veZero 
	\quad \forall \, \ve X \in \omref
	\quad \glmwith \quad 
	\te P \cdot \ve N = \hat{\ve t} \quad \forall \, \ve X \in \domreft
	\label{eq:imp}
\end{equation}
{and the incompressibility condition}
\begin{equation}
	 J-1 = 0 \quad \forall \, \ve X \in \omref
	\point \label{eq:pu}
\end{equation}
Inserting \eqref{eq:pu} and $-\pp=p$ into the stress definition \eqref{eq:PK1}, the first Piola-Kirchhoff stress can be rewritten as 
\begin{equation}
	\te P = \diffp{\psi}{\te F} 
	= \diffp{\psii(\fiso)}{\te F} - p \, \te F^{-\top}
	= \pki  - p \, \te F^{-\top}
	\label{eq:PK1-red}
	\commaf
\end{equation} 
wherein $p$ is the hydrostatic pressure and the stress contribution arising from isochoric deformation is assigned the symbol $\pki$.
Provided that the potential $\psii$ is objective, compatibility with the balance of angular momentum
\(
	\tg \upsigma = \tg \upsigma^\top
	\Leftrightarrow
	\te P \cdot \te F^{\top} = \te F \cdot \te P^{\top}
\)
is automatically fulfilled, see~\cite[Proposition~8.3.2]{silhavy1997}.
Moreover, evaluating the Clausius-Planck inequality
\(
	\te P \, : \dot{\te F}^\top - \dot \psi \geq 0 
\)
following~\cite{coleman1967}, consistency of the model with the Second law of thermodynamics can be proven.

\section{Admissible invariants and specific deformation states}
\label{sec:adm-inv}

The computation of the invariants $\ici, \iici$ for a given deformation state, i.e. given $\fiso$ or $\ciso$ is straightforward.
However, it may also be of interest which values of $\ici$ and $\iici$ are actually physically admissible. In other words, we aim at investigating which tuples $\is$ could arise from physically admissible deformations in a specimen.
Apart from its relevance for planning experiments, this discussion of the admissibility of invariants does also play a role when investigating the mathematical properties of material models formulated in deformation invariants, e.g. non-negativity of the free energy as in~\cite[Sect.~3]{dammass2025}. 
%
In this Section, we first analyse which combinations of invariants $\is$ can be reached by physically admissible deformation states.
Subsequently, we demonstrate that simple loading scenarios correspond to limiting cases in terms of the invariants.

\subsection{Set of admissible invariants}
\begin{figure}[bp!]
	\centering
	\includegraphics[width=0.67\linewidth,trim=0mm 0mm 0mm 0mm]{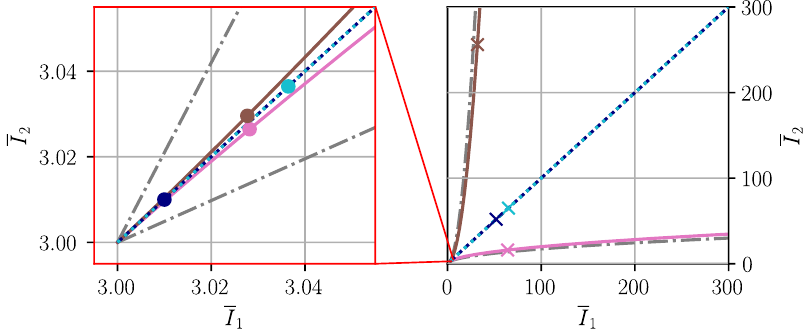} 
	\includegraphics[scale=0.8,trim=0mm 2mm 5mm -2mm]{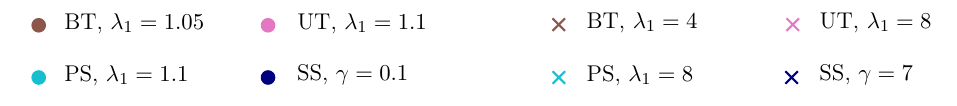}		
	\includegraphics[scale=0.8,trim=0mm 5mm 15mm 0mm]{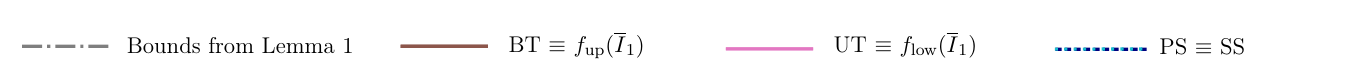}		
	\caption{Isochoric deformation invariants $\ici$, $\iici$: Bounds of the set of admissible invariant tuples $A$: Estimate from Lemma~\ref{lem:InvEstim} and sharp bounds $f_\nv{up}$ and $f_\nv{low}$ from Theorem~\ref{theo:InvSharp}.
		In addition, values for the scenarios of \textit{Uniaxial tension (UT)}, \textit{Equi-biaxial tension (BT)}, \textit{Pure shear (PS)} and \textit{Simple shear (SS)} for small (zoom on the left) or more significant (right) amounts of deformation are provided. Note that the sharp bounds $f_\nv{up}$ and $f_\nv{low}$ correspond to UT and BT.
	}
	\label{fig:admInv}
\end{figure}
We aim at specifying the set of all physically admissible tuples of invariants $\is$. In what follows, this set will be denoted as $A \subset \rset^2$.
We start our derivations by a definition of admissibility inspired by~\cite[Definition~C.1]{linden2023}:
\begin{definition}
	Let $\left(\ici, \iici \right) \in \rset^2$. $\left(\ici, \iici \right)$ is called an admissible invariant tuple, if and only if there exists a unimodular second order tensor $\ciso$ with principal invariants $\ici$, $\iici$ and positive real eigenvalues. 
\end{definition}
In what follows, we let $A$ denote the set of all admissible invariant tuples. A first estimate for $A$ can be established based on the following necessary condition for admissibility as provided in~\cite{moriarty1971,schroeder2003,hartmann2003b}:
\begin{lemma} Let the invariant tuple $\left(\ici, \iici \right) \in \rset^2$ admissible. Then it holds
	\begin{equation}
		 \left(\ici, \iici \right) \in B= \left\{ \left(\ici, \iici \right)\in [3,\infty) \times [3,\infty) \left|  \sqrt{3 \, \ici } \leq \iici \leq {\overline I}^2_1/3\right.\right\}
		\point
	\end{equation}
	\label{lem:InvEstim}
\end{lemma}
\begin{proof}
	From Hartmann and Neff, \cite[Corollary~A.3]{hartmann2003b}, we have $\ici \geq 3$ and $\iici \geq 3$ as well as $\iici \leq {\overline I}^2_1/3$. Moreover, \cite[Corollary~A.2]{hartmann2003b} yields $\sqrt{3 \, \ici } \leq \iici$.
\end{proof}
However, the estimate of Lemma~\ref{lem:InvEstim} gives a necessary condition, which is not sufficient: There are tuples $\ve t \in B$, which do not correspond to admissible deformation states, i.e. $\ve t \notin A$, cf. Fig.~\ref{fig:admInv}.
Nevertheless, with Lemma~\ref{lem:InvEstim} at hand, a precise definition of $A$ as a semi-infinite subset of $\rset^2$ can be delivered:
\begin{theorem}
	Let $\left(\ici, \iici \right) \in \rset^2$. $\left(\ici, \iici \right)$ is an admissible invariant tuple if and only if
	\begin{equation}
		\left(\ici, \iici \right) \in A = \left\{ \left(\ici, \iici \right)\in [3,\infty) \times [3,\infty) \left|  f_\nv{low} (\ici) \leq \iici \leq f_\nv{up}(\ici)\right.\right\}
		\comma
	\end{equation}
	with the upper and lower bounds for $\iici$ for a given $\ici$ expressed by
	\begin{align*}
		f_\nv{low} \left(\ici\right) =& 
		-\frac{1}{24} \,\sqrt [3]{\ici^{6}-540\,\ici^{3}-5832+24\,\sqrt {-3\,\ici^{9}+243
				\,\ici^{6}-6561\,\ici^{3}+59049}} \\
			&+{\frac {6\, \left( -3/2\,\ici-{{\ici^{4}}/{144}} \right) }
				{\sqrt [3]{\ici^{6}-540\, \ici^{3}-5832+24
				\,\sqrt {-3\,\ici^{9}+243\,\ici^{6}-6561\,\ici^{3}+59049}}}
			} +\frac{1}{12} \,\ici^{2}\\
			&+\frac{i}{2} \sqrt {3}
			\Bigg(
			 \frac{1}{12} \,\sqrt [3]{\ici^{6}-540\,\ici^{3}-5832+24\,\sqrt {
				-3\,\ici^{9}+243\,\ici^{6}-6561\,\ici^{3}+59049}} \\
			&\quad +  12\,{\frac {1}{\sqrt [3
				]{\ici^{6}-540\,\ici^{3}-5832+24\,\sqrt {-3\,\ici^{9}+243\,\ici^{6}-6561\,
						\ici^{3}+59049}}} \left[ - \frac{3}{2}  \, \ici -{\frac {\ici^{4}}{144}} \right] }
			\Bigg) 
%
%
		\numberthis \label{eq:flow} \\
		\intertext{and}
		f_\nv{up}\left(\ici\right) =& 
		\frac{1}{12}\,\sqrt [3]{\ici^{6}-540\,\ici^{3}-5832+24\,\sqrt {-3\,\ici^{9}+243\,
				\ici^{6}-6561\,\ici^{3}+59049}} \\
			&-\,{\frac {12 \, \left( -3/2\, \ici -\ici^{4}/{144} \right)}
				{\sqrt [3]{\ici^{6}-540\,\ici^{3}-5832+24\,\sqrt {-3\,\ici^{9}+243\,\ici^{6}-6561\,\ici^{3}+59049}}}
			 }+\frac{1}{12}\,\ici^{2}
		%
		\comma
		\numberthis \label{eq:fup}
	\end{align*}
    wherein $i$ is the imaginary unit.%
	\footnote{Note that $f_\nv{low}(\ici)$ is real-valued for $\ici \in [3,\infty)$ by Cardano's rule, even though the imaginary unit $i$ is used in its definition.}
	\label{theo:InvSharp}
\end{theorem} 
\begin{proof}
	Following Cardano's rule, see e.g. \cite[Sect.~1.6.2.3]{bronshtein2015}, the characteristic equation \eqref{eq:charpol} has three real solutions if and only if the discriminant
	\begin{align*}
		D(\ici,\iici) & = \left(\frac{1}{3} \, \iici - \frac{1}{9} \, \ici^2\right)^3
		+ \left(\frac{-\ici^3}{27} + \frac{\ici \, \iici}{6} - \frac{1}{2}\right)^2 \numberthis \\
		&= \frac{1}{27} \,\left( \ici^3 + \iici^3 - \frac{1}{4} \, \ici^2 \, \iici^2 - \frac{9}{2} \, \ici \, \iici + \frac{27}{4} \right)
	\end{align*}
	is non-positive. Accordingly, only such $(\ici,\iici) \in B$ are admissible for which $D(\ici,\iici) \leq 0$ holds. 
	In order to find the subset of $B$, for which $D\leq 0$, we study the case $D(\ici,\iici)=0$, which is a cubic equation in both $\ici$ and $\iici$ and provide explicit representations of $\iici(\ici)$ for this case.
	For $(\ici,\iici) \in B$, $D(\ici,\iici)=0$ has only real solutions for $\iici$, since the corresponding discriminant
	\begin{equation}
		E(\ici) = -\frac{\left(\ici-3\right)^3 \, \left(\ici^2+3\, \ici +9\right)^3}{1728}
	\end{equation} 
	is non-positive for $\ici \in [3,\infty)$.
	Again employing Cardano's rule yields
	\begin{equation}
		D(\ici,\iici) = 0
		\Longleftrightarrow
		\iici \in \left\{f_\nv{out}(\ici),f_\nv{low}(\ici),f_\nv{up}(\ici)\right\}
		\label{eq:charpolZeroSol}
	\end{equation}
	with $f_\nv{low}(\ici)$, $f_\nv{up}(\ici)$ as given in \eqref{eq:flow} and $f_\nv{out}(\ici)$ as given in \eqref{eq:fout} in App.~\ref{sec:explSolDZero}.
	Since $f_\nv{out}(\ici) < 3$, its graph does not lie in $B$ and therefore it does not give any further insight due to Lemma~\ref{lem:InvEstim}. In contrast, the graphs of $f_\nv{low}(\ici)$ and $f_\nv{up}(\ici)$ do both lie in $B$.
	Moreover, it can be easily verified that
	\begin{align*}
		&\exists \, \left(\ici, \iici \right) \in \tilde A = \left\{ \left(\ici, \iici \right)\in (3,\infty) \times (3,\infty) \left|  f_\nv{low} (\ici) < \iici < f_\nv{up}(\ici)\right.\right\} :
		D(\ici,\iici)< 0  \\
		&\Longleftrightarrow D\left(\ici, \iici \right)< 0 \quad \forall \, (\ici,\iici) \in \tilde A \comma \numberthis
	\end{align*}
	such that there are three real eigenvalues if and only if $\left(\ici, \iici \right) \in A$.
	Finally, we complete the proof by inserting Lemma~\ref{lem:posEV}, which states positivity of the real solutions of the characteristic equation.
\end{proof}
Together with the estimate $B$ from Lemma~\ref{lem:InvEstim}, the set of admissible invariants $A$ is visualised in Fig.~\ref{fig:admInv}.
Moreover, tuples $\is$ are provided for several simple deformation states at small and more significant amounts of deformation.
From this visualisation in Fig.~\ref{fig:admInv}, it can also be concluded that only a very small subset of $A$ is relevant in the setting of small strains.%
\footnote{After publication of the final version of the manuscript, Python code for reproducing Fig.~\ref{fig:admInv} will be available at \url{https://github.com/NEFM-TUDresden}.}

\subsection{Invariants for specific load cases}
\label{sec:specLC}

For experimental investigations, several simple deformation states may be of particular interest, where the stretch $\lambda_1 \in \rsp$ or the amount of shear $\gamma \in \rset$ is prescribed as follows:
\begin{enumerate}[label=(\alph*)]
	\item \textit{Uniaxial tension} (\textit{UT;} $\lambda_1>1$) or \textit{compression} (\textit{UC;} $\lambda_1<1$):
	\[\te F \equiv \fiso = \diag(\lambda_1,\lambda_1^{-1/2},\lambda_1^{-1/2})
	\quad \Longrightarrow \quad
	\ici = \lambda_1^2 + 2 \, \lambda_1^{-1}, \quad
	\iici = \lambda_1^{-2} + 2 \, \lambda_1
	\]
	\item \textit{Equi-biaxial tension} (\textit{BT;} $\lambda_1>1$) or \textit{compression} (\textit{BC;} $\lambda_1<1$): 
	\[
	\te F \equiv \fiso = \diag(\lambda_1,\lambda_1,\lambda_1^{-2})
	\quad \Longrightarrow \quad
	\ici = 2 \, \lambda_1^2 + \lambda_1^{-4}, \quad
	\iici = 2\, \lambda_1^{-2} + \lambda_1^4
	\]
	\item \textit{Pure shear (PS):} \[\te F \equiv \fiso = \diag(\lambda_1,1,\lambda_1^{-1})
	\quad \Longrightarrow \quad
	\ici = \iici = 1+ \lambda_1^2 + \lambda_1^{-2} \]
	\item \textit{Simple shear (SS):} 
	\[[F_{kL}] \equiv [\overline{F}_{kL}] = 
	\begin{bmatrix}
		1 & \gamma & 0\\ 0 & 1 & 0 \\ 0 & 0 & 1
	\end{bmatrix}
	\quad \Longrightarrow \quad
	\ici = \iici = 3 + \gamma^2
	\]
\end{enumerate}
For both practical and theoretical reasons, these deformation states are appealing:
On the one hand, the hydrostatic pressure may be identified from the boundary condition \eqref{eq:imp}\textsubscript{2} in these cases. This allows to identify the stress contribution from isochoric deformations $\pki$ directly. Moreover, several benchmark data sets for soft matter considering these scenarios are available in the literature, see Sect.~\ref{sec:fits}.
On the other hand, these deformation states correspond to special cases in terms of the invariants, as shown in Fig.~\ref{fig:admInv}.
Firstly, $\ici$ and $\iici$ are identical for both {PS} and {SS}.%
\footnote{The question where the denomination \textit{shear} in PS and SS stems from is not trivial. For details, the reader is referred to \cite{thiel2019}.}
Secondly, {UT}/UC and {BT}/BC can be shown to correspond to limiting deformation states in terms of the invariants:
\begin{corollary}
	Let $\is$ correspond to UT/UC and BT/BC. Then $\is$ lies on the boundary of the set of admissible invariants $A$.
\end{corollary}
\begin{proof}
	For UT/UC and BT/BC, at least two eigenvalues of $\ciso$ are identical. Therefore, $D\is = 0$ holds for the discriminant of the characteristic equation by Cardano's rule.
\end{proof}
As a consequence, the explicit representations for $\iici$  as a function of $\ici$ as provided above are valid for UT and BT: $\iici = f_\nv{low}(\ici)$ for UT and $\iici = f_\nv{up}(\ici)$ for BT, when $\lambda_1 \geq 1$, and vice versa $\iici = f_\nv{up}(\ici)$ for UC and  $\iici = f_\nv{low}(\ici)$ for BC, when $\lambda_1 \leq 1$.

In addition, note that all the deformation states UT/UC, BT/BC, PS and SS appear in the $\ici$-$\iici$ diagram as lines that are straight to an excellent degree of approximation, if only small deformations are considered, see Fig.~\ref{fig:admInv}.
Moreover, there are only small differences between the slope of these lines between the individual load cases. In other words, there is only minor difference between different deformation types in the small strain regime in terms of~\mbox{$(\ici, \iici)$}, whereas the distance between invariant tuples corresponding to different deformation types is way more pronounced for larger amounts of deformation.

\section{When and why $\ici$ and $\iici$ are important for precise models of hyperelasticity}
\label{sec:fits}

Above in Sect.~\ref{sec:adm-inv}, we have shown that the physically admissible incompressible deformation invariants form a semi-infinite surface in $\rset^2$. Therefore, from a strictly kinematical point of view it is not suitable to consider only one invariant in a material model, which is intended to be reliable for arbitrary isochoric deformations.
However, many classical models are solely based on $\ici$, see e.g. \cite[Tab.~7--8]{ricker2023}, and recently, even an NN-based model considering only $\iici$ has been proposed~\cite{kuhl2024}. 
In this Section, we investigate the performance of models considering solely either $\ici$ or $\iici$, and compare the performance to a formulation depending on both $\ici$ and $\iici$.
To this end, we study three different experimental benchmark data sets from the literature, which do all provide results for multiple deformation scenarios: Treloar's rubber data~\cite{treloar1944}, the pseudo-elastic response of human cortex matter of Budday~et~al.~\cite{budday2017}, and results for neoprene of Alexander~\cite{alexander1968}.
As a model ansatz, we exemplarily consider a Physics-augmented neural network {(PANN)} for the isochoric strain energy potential $\psii$~\cite{dammass2025}, since PANNs have been shown to be flexible and precise ansatzes for various materials~\cite{klein2022,klein2022a,kalina2023,linden2023,kalina2024,rosenkranz2024b,fuhg2024d}.
Note that other invariant-based data-driven modelling approaches, e.g. Constitutive artificial neural networks (CANNs)~\cite{linka2023} or spline-based approaches~\cite{wiesheier2024} may also be considered.
For the scope of this work, we focus on the role of the invariants serving as model input rather than on the specific model architecture, such that the key findings provided in the following depend only very little on the choice of the particular model ansatz.

An implementation of all the models considered in this work and the specific parameters for the examples discussed below are provided as supplementary material.%
\footnote{After publication of the final version of the manuscript, the code will be available at \url{https://github.com/NEFM-TUDresden}.}

\subsection{Physics-augmented neural network material model}
\label{sec:pann}

Within this work, we consider an invariant-based PANN potential for the incompressible material response, which has been introduced in our previous work~\cite{dammass2025}.
In the following, we briefly summarise the architecture of the PANN and the strategy for its parameterisation.

\subsubsection{Neural network architecture}
\label{sect:nn_archi}

\begin{figure}[bp!]
	\centering
	\includegraphics[width=0.9\linewidth,trim=0mm 0mm 0mm 0mm]{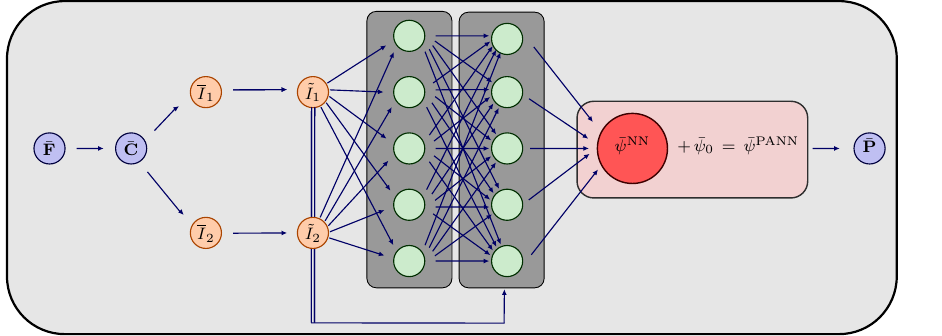} 
	\caption{Schematic overview on the architecture of the isochoric PANN potential $\psiinn$}
	\label{fig:pann_iso}
\end{figure}

The PANN model $\psiinn$ is constructed such that it a priori fulfils the desirable properties of hyperelastic potentials. For the case of an incompressible and isotropic material, these can be specified as follows~\cite{dammass2025,linden2023}:
\begin{enumerate}[label=(\roman*)]
	\item {Thermodynamic consistency:}
	\( \pki = \dfrac{\partial \psii(\fiso)}{\partial \te F} \)
	\item {Objectivity:}
	$\psiinn(\te Q \cdot \fiso) =  \psiinn(\fiso) \,\, \forall \, \te Q \in \so$
	\item {Isotropy:}
	$\psiinn(\fiso \cdot \te Q) =  \psiinn(\fiso) \,\, \forall \,\te Q \in \so$
	%
	%
	\item {Non-negativity:} 
	$\psiinn(\fiso) \geq 0 \,\, \forall \, \fiso \in \tesu$
	\item {Zero energy in the undeformed state:}
	$\psiinn(\fiso = \te I) = 0$
	\item {Zero stress in the undeformed state:}
	$\pki(\fiso = \te I) = \te 0$
	\item {Polyconvexity:}%
	\footnote{Polyconvexity can be considered an important property of hyperelastic potentials and is of particular relevance for proving the existence of energy minimisers~\cite{ball1976}. 
		Moreover, polyconvexity of the potential implies ellipticity of the associated Euler equations, which can be interpreted as material stability \cite{zee1983}, see also \cite{bonet2015}.
		In addition, it has been shown that polyconvexity can significantly improve the capability of models to extrapolate compared to non-polyconvex alternatives~\cite{linden2023}.
		However, it seems worth mentioning that a non-polyconvex potential does neither {a priori} violate  physical laws nor {a priori} predict unrealistic material behaviour. On the contrary, especially in the context of multiscale modelling and computational homogenisation, enforcing global polyconvexity can hinder model accuracy, see \cite{kalina2024}.}
	There is an $\mathcal F(\te F, \cof \te F, J) = \psii(\fiso)$ that is convex w.r.t. $(\te F, \cof \te F, J)$
\end{enumerate}
In order to enable polyconvexity of $\psiinn$, the polyconvex invariants 
\begin{equation}
	\isc = \left( \ipci, \ipcii\right) = \left( \ici, \iici^{3/2}\right) \comma
\end{equation}
see~\cite[Lemma~2.2]{hartmann2003b}, are considered as input of the actual neural network $\psiinon$, since  $\iici$, different from $\ipcii$, is not polyconvex \cite[Lemma~2.4]{hartmann2003b}.
For $\psiinon$, following~\cite{amos2017} and later~\cite{linden2023}, we adopt a Fully input convex neural network~{(FICNN)} with skip connections, which is non-decreasing. 
For this architecture and the general case of $H \in \mathbb N_{>0}$ hidden layers with each hidden layer $h \in \left\{1, \dots, H\right\}$ possessing $N^{\text{N},h} \in \mathbb N_{>0}$ neurons, the network can be specified as%
\footnote{Adopted from \cite{kalina2024} and as in \cite{dammass2025}, non-trainable normalisation layers are used that perform a simple linear scaling of the input and output quantities for numerical purpose. Details are omitted here, for the reason of brevity, and the reader is referred to \cite[App.~D]{kalina2024}.}
\begin{align}
	&\psiinon\left(\isc\left(\fiso\right)\right)
	= \sum_{\alpha=1}^{N^{\text{N},H}} W_{\alpha}\, f^{H}_\alpha + 
	\sum_{\beta=1}^2 S_{\beta} {\tilde I}_\beta + B \commaf \label{eq:ICNN_out}\\
	%
	\intertext{with the output of the neurons of the $h$-th hidden layer for $h \in \left\{2, \dots, H\right\}$ given by}
	&f^{h}_\alpha =  
	\afu
	\left[\sum_{\beta=1}^{N^{\text{N},h-1}}w_{\alpha\beta}^{h} \, f_\gamma^{h-1} +
	\sum_{\beta=1}^2 s_{\alpha\beta}^{h} \, {\tilde I}_\beta + b_\alpha^{h} \right]
	\quad \glmwith \quad \alpha \in \left\{1, \dots, N^{\text{N},h}\right\} \\
	%
	\intertext{and the output of the neurons of the first hidden layer ($h=1$) by}
	&f^{1}_\alpha =  
	\afu \left[\sum_{\beta=1}^2 w_{\alpha\beta}^{1} \, {\tilde I}_\beta + b_\alpha^{1}\right]
	\quad \glmwith \quad \alpha \in \left\{1, \dots, N^\text{N,1}\right\}
	\point
\label{eq:ICNN_first}
\end{align}
We consider the convex and non-decreasing \textit{softplus} activation function
\begin{equation}
	\afu : \rs \longrightarrow \rsp \quad,\quad x \longmapsto \ln \left[1 + \exp(x) \right]
	\comma
\end{equation}
and the non-negative weights $W_\alpha$, $S_\beta$,  $w_{\alpha \beta}$ and $s_{\alpha\beta}^h$ as well as the bias values $b_\alpha^h$ and $B$.
For the study on the performance of the model considering solely one invariant, all weights with respect to the other are set to zero, for instance for the $\ipci$-only model $w_{\alpha 2}^{h} = s_{\alpha 2}^{h} = 0 \, \forall \, h, \alpha $  as well as $S_2 =0$.
With this definition of $\psiinon$ at hand, the architecture of the PANN is specified as 
\begin{equation}
	\psiinn\left(\isc(\fiso)\right) = \psiinon\left(\isc(\fiso)\right) + \psiinno \comma
\end{equation}
wherein the constant
\begin{equation}
	\psiinno = - \psiinon\left(\isc(\fiso = \te I)\right) \in \rs
\end{equation}
is included in order to ensure zero strain energy in the undeformed state.
This architecture of the PANN is visualised in Fig.~\ref{fig:pann_iso}.
For comparison, we also consider unconstrained feedforward neural networks (FNN) in the numerical examples below. These are obtained from the architecture specified in \eqref{eq:ICNN_out}--\eqref{eq:ICNN_first} by allowing negative weights $W_\alpha$, $w_{\alpha \beta}$ and discarding the skip connections, i.e. $S_\beta = 0$, $s_{\alpha\beta}^h=0$.
In all the numerical examples shown below, simple networks consisting of two hidden layers with up to four neurons each are considered for the respective PANN or FNN, since performance could not be remarkably improved with a more complex architecture.
The specific parameters and implementations of the models are provided in the supplementary material.

\subsubsection{Training procedure}
\label{sec:pann-training}

For the three different benchmark data sets considered below, experimentally-determined stresses are available, for which the load cases UT, UC, BT, PS or SS (see Sect.~\ref{sec:specLC}) have been investigated.
Data for  $P_{11}$ is available for UT, UC, BT and PS, while the shear stress $P_{12}$ has been determined in the SS experiments.
In order to parameterise the neural networks from experimental data, a first order \textit{Sobolev training}~\cite{czarnecki2017,vlassis2020} is carried out.
The parameters of the PANN are identified from the optimisation problem
\begin{align*}
	&\left([W_\alpha], [S_\beta],B,[w_{\alpha \beta}^h], [s_{\alpha\beta}^h], [b_\alpha ]\right) =  \argmin_{ \left([W_\alpha], [S_\beta],B, [w_{\alpha \beta}^h], [s_{\alpha\beta}^h], [b_\alpha^h ]\right) \in Z}
	\mathcal L
	\\
	&\quad \glmwith \quad
	Z = \rsnn^{N^{\nv{N},H}} \times \rsnn^{2} \times \rset \times\rsnn^{2\times 2} \times \prod_{h =2}^H\rsnn^{N^{\nv{N},h-1} \times N^{\nv{N},h}} \times \prod_{h =2}^H\rsnn^{N^{\nv{N},h}\times 2} \times \prod_{h =1}^H\rset^{\nv{N},h}
	\numberthis
	\commaf
\end{align*}
which is solved numerically, and in the same manner, the parameters of the FNN are identified. To this end, the quasi Newton optimiser \textit{SLSQP (Sequential Least Squares Programming)} and the stochastic gradient-based optimiser \textit{Adam} have been used, cf.~\cite[Appendix~G]{kalina2024}.%
\footnote{As discussed in \cite{kalina2024}, the SLSQP optimiser is typically advantageous for the training of small and moderately-seized networks in terms of both performance of the fit and computational efficiency. 
For this work, the Adam and SLSQP have been compared for all numerical examples, and the observations of \cite{kalina2024} could be confirmed, in principle.
However, for some of the models considering only one invariant, Adam gave preferred results.}
The \textit{loss function} is defined as
\begin{align*}
	\mathcal L = & \,
	N^\nv{UT} \,
	\mse\left({}^\nv{PANN}\ve P^\nv{UT}, {}^\nv{ex}\ve P^\nv{UT} \right)
	+ 	N^\nv{UC} \,
	\mse\left({}^\nv{PANN}\ve P^\nv{UC}, {}^\nv{ex}\ve P^\nv{UC} \right)
	+   N^\nv{BT} \,
	\mse\left({}^\nv{PANN}\ve P^\nv{BT}, {}^\nv{ex}\ve P^\nv{BT} \right) \\
	& +   N^\nv{PS} \,
	\mse\left({}^\nv{PANN}\ve P^\nv{PS}, {}^\nv{ex}\ve P^\nv{PS} \right)
	+ N^\nv{SS} \,
	\mse\left({}^\nv{PANN}\ve P^\nv{SS}, {}^\nv{ex}\ve P^\nv{SS} \right)
	\numberthis
\end{align*}
with the vectors of experimentally-determined values of $P_{11}$ denoted as ${}^\nv{ex}\ve P^\nv{LC}$ and the according vector of values predicted by the model given by ${}^\nv{PANN}\ve P^\nv{LC}$ for each load case $\text{LC} \in \{\text{UT},\text{UC},\text{BT},\text{PS}\}$.
Likewise, the vectors of shear stresses $P_{12}$ are denoted as ${}^\nv{PANN}\ve P^\nv{SS}$ and ${}^\nv{ex}\ve P^\nv{SS}$.
The non-negative constants $N^\nv{UT} ,N^\nv{UC}, N^\nv{BT}, N^\nv{PS}, N^\nv{SS} \in \rs_{\geq 0}$ are incorporated in order to account for the variation in the number of experimentally investigated states between UT, UC, BT, PS and SS.%
\footnote{The case that no experimental data for the load case LC is available in one of the benchmarks  investigated below is recovered for $N^\nv{LC}=0$.}
Furthermore,
\begin{equation}
	\mse : 
	(\ve a,  \ve b) \longmapsto \frac{1}{n} \sum\limits_{i=1}^{n} (a_i-b_i)^2
\end{equation}
is the \textit{mean squared error} function.
In addition to $\mse$, the coefficient of determination
	\begin{equation}
		r^2 = 1 - \, \dfrac{\sum\limits_{i=1}^{n} \left(\prescript{\nv{ex}}{i}{P}-\prescript{\nv{PANN}}{i}{P}\right)^2}{\sum\limits_{i=1}^{n} \left(\prescript{\nv{ex}}{i}{P}- \prescript{\nv{ex}}{}{\check{P}}\right)^2} 
		\quad \glmwith \quad
		\prescript{\nv{ex}}{}{\check{P}} = \sum\limits_{i=1}^{n} \prescript{\nv{ex}}{i}{P}
		\commaf
	\end{equation}
is used for judging the performance of a parameterised model.%
\footnote{Note that the symbol $r^2$, which is used for agreement with the notation typically used in the literature, is a certain abuse of notation, since $r^2<0$ is possible in case of a modest agreement between experimental data and model, although $r^2$ does only depend on real quantities.}

Since incompressibility of the material is assumed, the hydrostatic pressure $p$ (and likewise $\pp=-p$) is not determined by $\psiinn$.
However, for UT/UC, BT/BC and PS, $p$ can be identified exploiting the sparsity of $\te P$ and~$\te F$. Specifically, the hydrostatic pressure can be computed from $P_{33}=0$ in these cases. The corresponding stress tensors can be assumed to read
\begin{equation}
	\te P^\nv{UT/UC} = \diag(P_{11}, 0, 0) \commam
	\te P^\nv{BT/BC} = \diag(P_{11}, P_{11}, 0) \commam
	\te P^\nv{PS} = \diag(P_{11}, P_{22}, 0) 
	\commaf
	\label{eq:PK1_trelLC}
\end{equation}
respectively, and $P_{11}$ can be rewritten making use of the plane stress condition $P_{33}=0$:
\begin{equation}
	0 = P_{33} =  \overline{P}_{33} + \pp \, \frac{1}{\lambda_3} 
	\quad \Longleftrightarrow \quad
	P_{11} = \overline{P}_{11} - \overline{P}_{33}  \, \frac{\lambda_3}{\lambda_1} 
	\glmwith
	\lambda_3 = 
	\left\{\begin{matrix} &\lambda_1^{-1/2} &, \, \nv{UT \,\, or \,\, UC} \\ 
		&\lambda_1^{-2} &, \,  \nv{BT \,\, or \,\, BC} \\ 
		&\lambda_1^{-1} &, \,  \nv{PS \qquad \quad}\end{matrix}\right.
	\label{eq:pk1_11_expl_lcs}
\end{equation}
Evaluating the definition of stress \eqref{eq:PK1-red} for these scenarios and inserting $\pki$ into \eqref{eq:pk1_11_expl_lcs}, the relevant stress coordinate $P_{11}$ can be written as
\begin{align}
	&P_{11}^\nv{UT/UC} = 2 \, \left[ \left(\lambda_1 - \lambda_1^{-2}\right) \, \diffp{\psii}{\ici}
	+ \left(1- \lambda_1^{-3}\right)  \, \diffp{\psii}{\iici}
	\right]
	\comma
	\label{eq:PK1_11_UT}
	\\
	&P_{11}^\nv{BT/BC} = 2 \, \left[ \left(\lambda_1 - \lambda_1^{-5}\right) \, \diffp{\psii}{\ici}
	+ \left(\lambda_1^3 - \lambda_1^{-3} \right)  \, \diffp{\psii}{\iici}
	\right]
	\comma
	\label{eq:PK1_11_BT}
	\\
	&P_{11}^\nv{PS} = 2 \, \left(\lambda_1 - \lambda_1^{-3}\right) \, \left(\diffp{\psii}{\ici}
	+ \diffp{\psii}{\iici}
	\right)
	\comma
	\label{eq:PK1_11_PS}
\end{align}
for which the derivatives of the invariants as provided in \eqref{eq:derivI1}, \eqref{eq:derivI2} in Sect.~\ref{sec:appPK1SS} have been used.
For SS, a similar sparsity of $\te P$ as in \eqref{eq:PK1_trelLC} can not be assumed.
However, in this case, only the experimentally-determined shear stress $P_{12}$ is of interest in the scope of this work. For the invariant-based model considered here, the hydrostatic pressure does not give any contribution to this coordinate of the stress in case of SS deformation. In other words, $P_{12} = \overline{P}_{12}$ holds in case of SS, see Sect.~\ref{sec:appPK1SS} for all coordinates of the stress tensor. As a consequence, the model prediction of $\overline{P}_{12}$, which can be expanded as
\begin{equation}
	P_{12}^\nv{SS} = 2 \, \gamma\, \left(\diffp{\psii}{\ici}+ \diffp{\psii}{\iici} \right)
	\label{eq:PK1_12_SS}
\end{equation}
can be directly compared to the experimental values ${}^\nv{ex}P_{12}$.

\subsection{Treloar's data on natural rubber}
\label{sec:disc_trel}

\begin{figure}[tbp!]
	\centering
	\begin{subfigure}{\linewidth}
		\includegraphics[width=0.9\linewidth,trim=0mm 0mm 8mm 0mm]{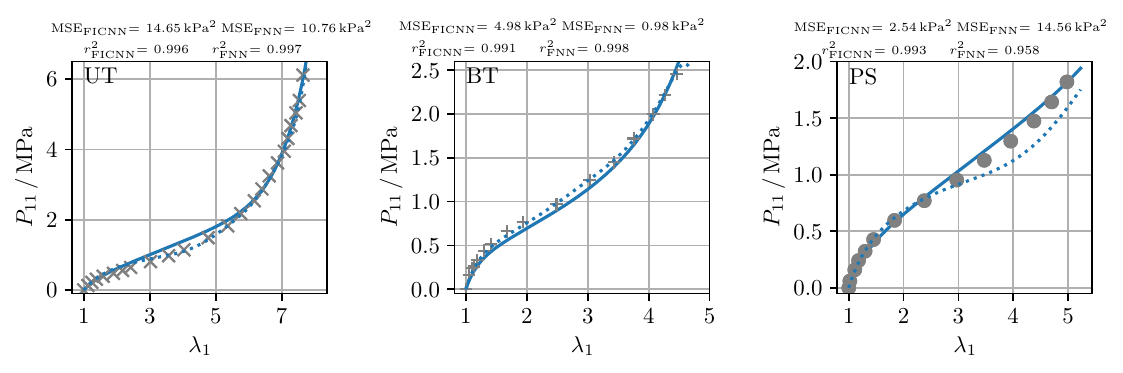} 
		\caption{Model including both $\ipci$ and $\ipcii$}		
		\label{fig:trell_all_both}
	\end{subfigure}
	\begin{subfigure}{\linewidth}
		\includegraphics[width=0.9\linewidth,trim=0mm 0mm 10mm 0mm]{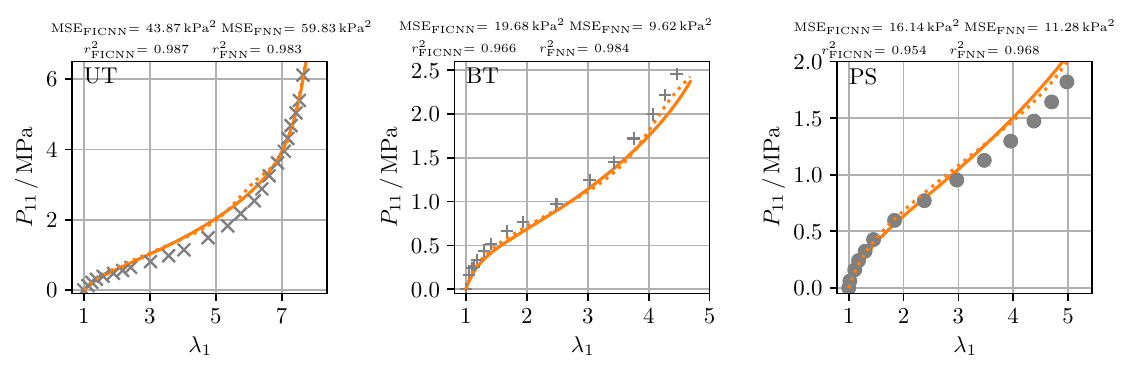} 
		\caption{Model including only $\ipci$. Note that the dotted lines overlap with the solid ones in a large stretch interval in all the subplots.}		
		\label{fig:trell_all_i1}
	\end{subfigure}
	\begin{subfigure}{\linewidth}
		\includegraphics[width=0.9\linewidth,trim=0mm 0mm 12mm 0mm]{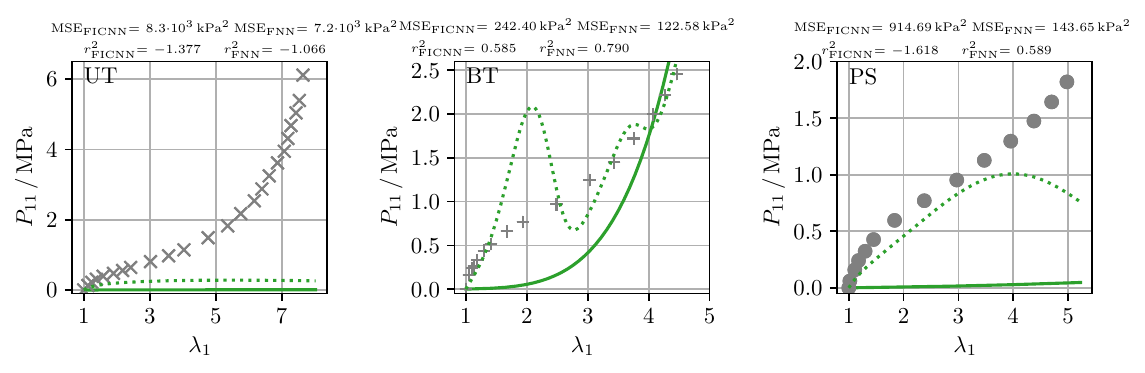} 
		\caption{Model including only $\ipcii$}		
		\label{fig:trell_all_i2}
	\end{subfigure}
	\includegraphics[scale=0.8,trim=0mm 0mm 5mm 0mm]{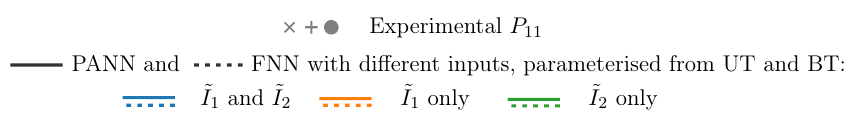}		
	\caption{Performance of the PANN (based on an FICNN) and an unconstrained FNN with respect to Treloar's rubber data~\cite{treloar1944} (\textit{NR-S} at a temperature of~$\SI{20}{\degree C}$) depending on the invariants considered as input: either both $\ipci$ and $\ipcii$~(a) or only $\ipci$~(b) or solely $\ipcii$~(c) are considered.
	Results for combined fit of data for uniaxial tension~(UT) and equi-biaxial tension~(BT) and the resulting prediction for pure shear~(PS) ($N^\nv{UT}=1$, $N^\nv{BT} = 3$, $N^\nv{PS} = 0$).
	}
	\label{fig:trel_fit_all}
\end{figure}
As a first benchmark, the data of Treloar~\cite{treloar1944} for \textit{NR-S}, i.e.~vulcanised natural rubber containing sulphur, at room temperature ($\SI{20}{\degree C}$) are considered.
These experimental results are widely used for benchmarking hyperelastic material models, see e.g.~\cite{steinmann2012,ricker2023,linka2023a,tac2024b,dammass2025}.%
\footnote{It seems worth mentioning that, despite Treloar's data is a well-established hyperelastic benchmark, \textit{NR-S} can show inelastic effects, for instance due to strain-induced crystallisation.}
In~\cite{treloar1944}, data is provided for UT, BT and PS. For the parameterisation, UT and BT are considered, while the stress responses to PS deformations are predictions for \textit{unseen} data ($N^\nv{UT}=1$, $N^\nv{BT} = 3$, $N^\nv{PS} = 0$).

The predictions of the parameterised PANNs and the experimental data are compared in Fig.~\ref{fig:trel_fit_all}.
As can be seen from Fig.~\ref{fig:trell_all_both}, very good agreement is reached for the architecture considering both $\ipci$ and $\ipcii$: for the coefficient of determination, $r^2 > 0.99$ is reached for UT and BT.
Moreover, this is also true for the prediction of PS, which has not been considered in training.
For comparison, an unconstrained FNN is also considered.
It is not surprising that the FNN can reach even slightly better agreement with the training data (UT and BT). However, prediction for the unseen PS load case is worse for the FNN than for the PANN.
The $\ipci$-only PANN, Fig.~\ref{fig:trell_all_i1}, does properly match the experiments in a qualitative manner. However, quantitative deviations are slightly more pronounced than for the $\ipci$ and $\ipcii$-model. 
This is in agreement with findings on different classical invariant-based hyperelastic models, see~\cite[Sects.~4.1.2 and 6]{ricker2023}.
Differences between PANN and FNN are less pronounced for the $\ipci$-only approach.
When the model is formulated solely depending on $\ipcii$, its behaviour is fundamentally different, see~Fig.~\ref{fig:trell_all_i2}.
The $\ipcii$-only PANN does not reflect any of the experiments adequately, nor does the unconstrained FNN. 
While the stress predicted by the PANN is almost zero regardless of $\lambda_1$ for UT and PS, it is at least in the same order of magnitude as the experimental values for BT. However, a convex dependency on $\lambda_1$ is predicted for BT for $\lambda_2 \gtrsim 2$, which is in disagreement with the experiment.
An explanation for these difficulties of the $\ipcii$-only formulation can be provided as follows. 
On the one hand, it has to be noted that $\ipcii$ increases very strongly with $\lambda_1$ for BT as shown in Fig.~\ref{fig:admInv}, which is not the case at all for UT and PS.
On the other hand, when $\ipci$-dependency of $\psii$ is dropped, the expressions for $P_{11}$ reduce to 
\begin{equation*}
	P_{11}^\nv{UT/UC} = 2 {\left(1- \lambda_1^{-3}\right)}  \, \diffp{\psii}{\iici}
	\comma \quad
	P_{11}^\nv{BT/BC} = 2 \, \left(\lambda_1^3 - \lambda_1^{-3} \right)  \, \diffp{\psii}{\iici}
	\quad \glmand \quad
	P_{11}^\nv{PS} = 2 \left(\lambda_1 - \lambda_1^{-3}\right) \, \diffp{\psii}{\iici}
	\point
\end{equation*}
In principle, these expressions for $P_{11}$ are of similar structure for all load cases. However, the significant dissimilarities in the first factors, e.g. $\left(1- \lambda_1^{-3}\right)$ (UT) vs. $\left(\lambda_1^3 - \lambda_1^{-3} \right)$ (BT),
do obviously hinder an agreement with experimental data for all load cases at the same time.

\begin{figure}[tbp!]
	\centering
	\begin{subfigure}{\linewidth}
		\includegraphics[width=0.9\linewidth,trim=0mm 0mm 9mm 0mm]{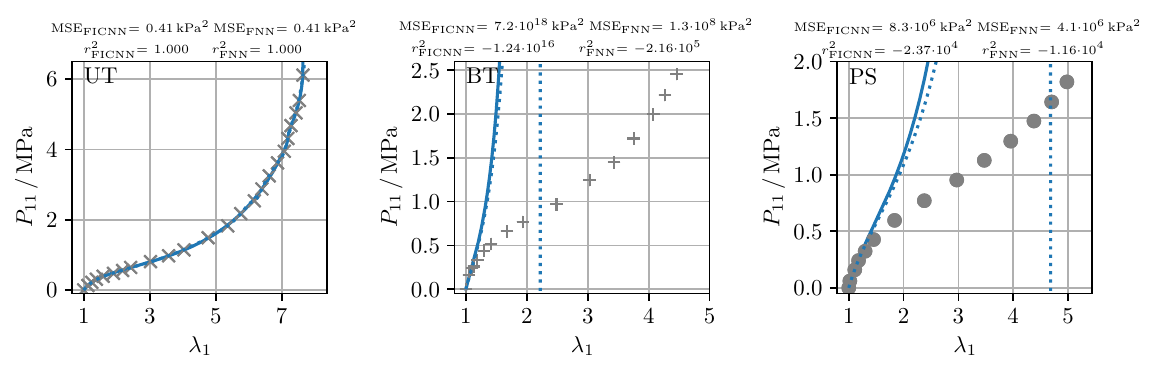} 
		\caption{Model including both $\ipci$ and $\ipcii$}		
		\label{fig:trel_ut_both}
	\end{subfigure}
	\begin{subfigure}{\linewidth}
		\includegraphics[width=0.9\linewidth,trim=-1mm 0mm 4mm 0mm]{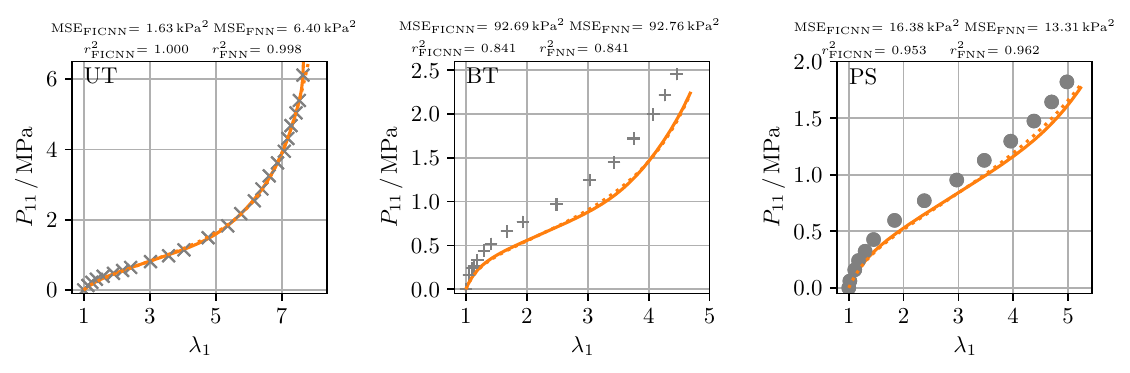} 
		\caption{Model including only $\ipci$}		
		\label{fig:trel_ut_i1}
	\end{subfigure}
	\begin{subfigure}{\linewidth}
		\includegraphics[width=0.9\linewidth,trim=0mm 0mm 12mm 0mm]{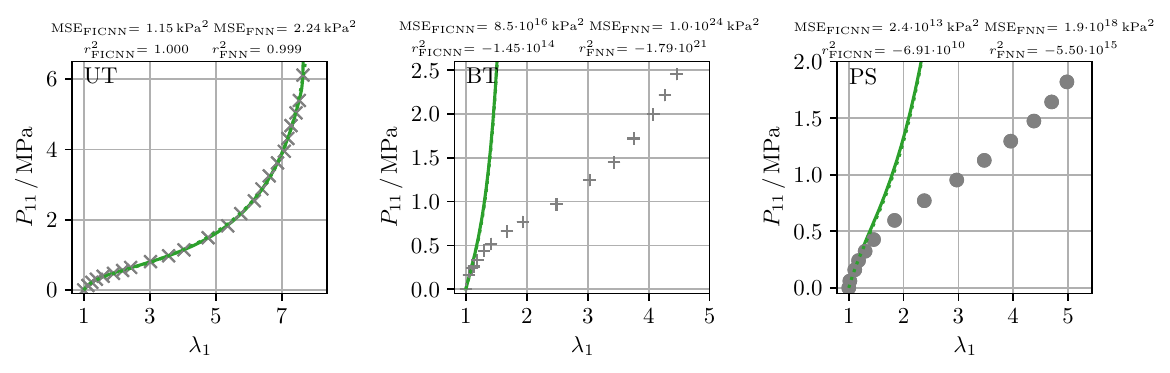} 
		\caption{Model including only $\ipcii$}		
		\label{fig:trel_ut_i2}
	\end{subfigure}
	\includegraphics[scale=0.8,trim=0mm 0mm 5mm 0mm]{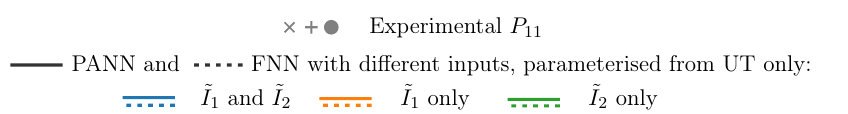}		
	\caption{Performance of the PANN (based on an FICNN) and an unconstrained FNN with respect to Treloar's rubber data~\cite{treloar1944} (\textit{NR-S} at a temperature of~$\SI{20}{\degree C}$) depending on the invariants considered as input: either both $\ipci$ and $\ipcii$~(a) or only $\ipci$~(b) or solely $\ipcii$~(c) are considered.
		Results for fit considering only uniaxial tension~(UT), while both equi-biaxial tension~(BT) and pure shear~(PS) are predictions to \textit{unseen} deformation states ($N^\nv{UT}=1$, $N^\nv{BT} = N^\nv{PS} = 0$).
		Note that the dotted lines overlap with the solid ones in a large stretch interval.}
	\label{fig:trel_fit_UT}
\end{figure}	
For further investigation, a parameterisation of the three model variants solely from UT ($N^\nv{UT}=1$,\linebreak \mbox{$N^\nv{BT} = N^\nv{PS} = 0$)} is shown in Fig.~\ref{fig:trel_fit_UT}.
In this case, regardless of considering both $\ipci$ and $\ipcii$ or only one of the invariants as input of the model, the response to UT can be depicted very precisely. However, for the $\ipcii$-only case, the stress under BT and PS is massively overestimated as shown in Fig.~\ref{fig:trel_ut_i2}.
The same is true for the model based on both $\ipci$ and $\ipcii$, Fig.~\ref{fig:trel_ut_both}, for which such a phenomenon has already been reported in the literature~\cite{dammass2025,ricker2023,boyce2000}: According to \cite{dammass2025,boyce2000} models including $\ipcii$-dependency should be parameterised based upon multiaxial deformation states in order to avoid overly stiff predictions.
An explanation of this phenomenon can be provided by means of Sect.~\ref{sec:adm-inv}, and descriptively by Fig.~\ref{fig:admInv}: If experimental data for model calibration is only available for one line within the semi-infinite surface of admissible invariants $A$, it is not possible to identify the parameters that quantify how sensitive the material response is regarding deviations away from this line.
In this context also note that $\iici=f_\nv{low}(\ici)$ for UT, and therefore $\ipcii$~takes the smallest admissible values for a given~$\ipci$ in case of UT.
Different from that, qualitatively correct predictions are also possible for BT and PS if only $\ipci$ is considered in the model, Fig.~\ref{fig:trel_ut_i1}.
Overall, the responses are very similar between PANN and FNN in all cases when only UT is considered for calibration of the models.

It can be concluded that $\ipci$ is strictly necessary for models of rubber elasticity. Considering $\ipcii$ enables more precision, yet makes it essential to evaluate multi-axial data for the parameterisation in order to avoid unphysical predictions for arbitrary loadings.
Moreover, an unconstrained FNN can enable better agreement with training data than the polyconvex PANN, yet may be less reliable in prediction of unseen data, see also~\cite{linden2023}.

\subsection{Human cortex benchmark data of Budday et al.}
\begin{figure}[tbp!]
	\centering
	\begin{subfigure}{\linewidth}
		\includegraphics[width=0.9\linewidth,trim=0mm 0mm 6mm 0mm]{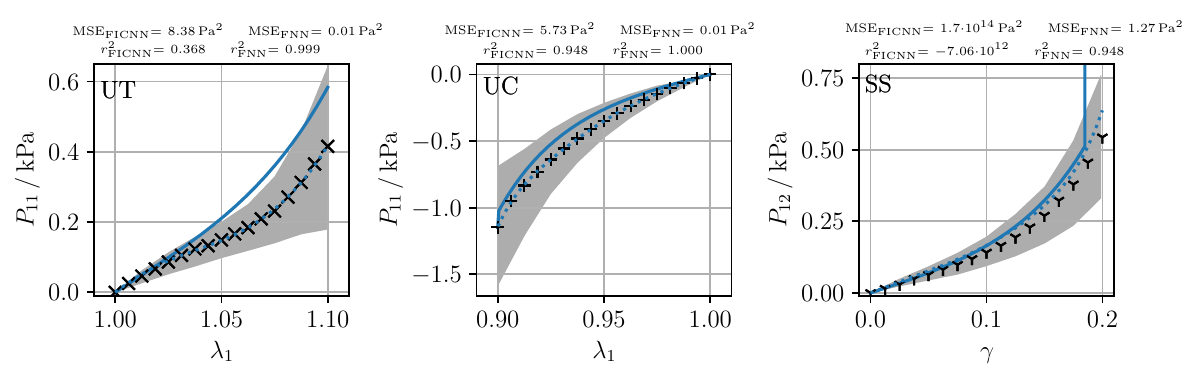} 
		\caption{Model including both $\ipci$ and $\ipcii$}		
		\label{fig:cortex_both}
	\end{subfigure}
	\begin{subfigure}{\linewidth}
		\includegraphics[width=0.9\linewidth,trim=0mm 0mm 6mm 0mm]{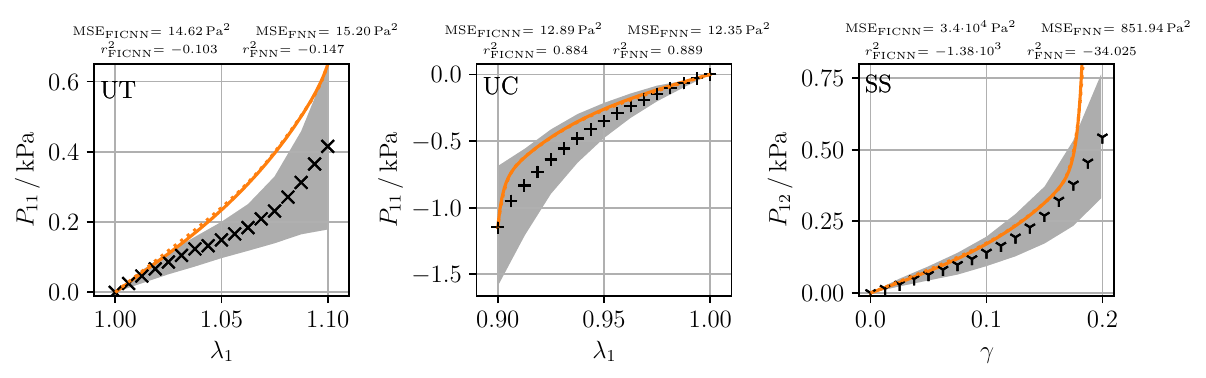} 
		\caption{Model including only $\ipci$. Note that the dotted lines overlap with the solid ones in a large stretch interval.}		
		\label{fig:cortex_i1}
	\end{subfigure}
	\begin{subfigure}{\linewidth}
		\includegraphics[width=0.9\linewidth,trim=0mm 0mm 0mm 0mm]{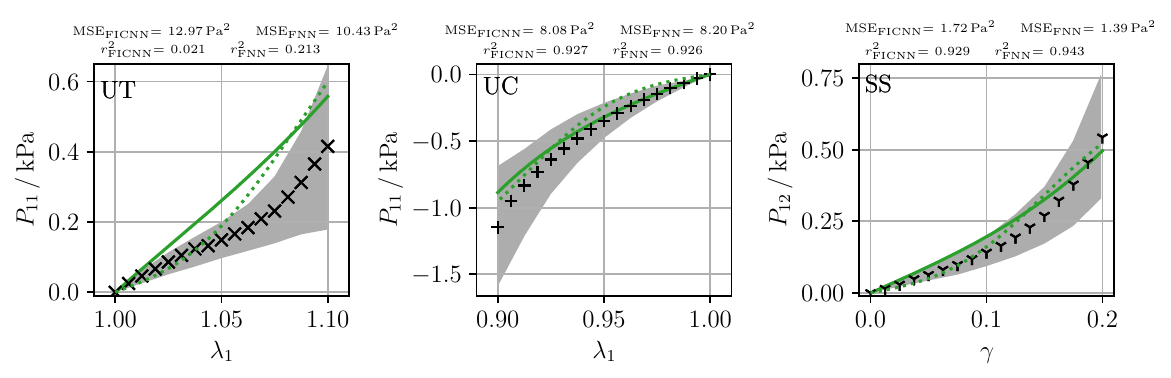} 
		\caption{Model including only $\ipcii$}		
		\label{fig:cortex_i2}
	\end{subfigure}
	\includegraphics[scale=0.8,trim=0mm 0mm 5mm 0mm]{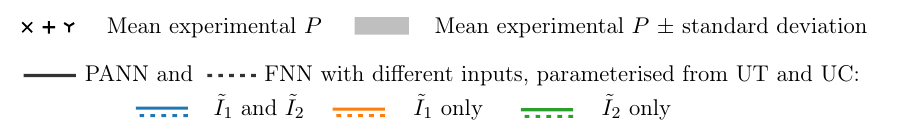}		
	\caption{Performance of the PANN (based on an FICNN) and an unconstrained FNN with respect to the human cortex benchmark data~\cite{budday2017} depending on the invariants considered as input: either both $\ipci$ and $\ipcii$~(a) or only $\ipci$~(b) or solely $\ipcii$~(c) are considered.
	Results for combined fit of pseudo-elastic uniaxial tension~(UT) and uniaxial compression~(UC), while the predicted responses to simple shear~(SS) are \textit{unseen} during training ($N^\nv{UT} = N^\nv{UC} = 1$, $N^\nv{SS}=0$).
	}
	\label{fig:cortex_fit}
\end{figure}

As second example, we consider the pseudo-elastic response of human cortex matter as investigated by Budday~et~al.~\cite{budday2017},
which has been frequently considered for benchmarking data-driven material models in recent years~\cite{linka2023,mcculloch2024,kuhl2024,abdolazizi2025}.
\footnote{Note that, although the dataset on human cortex matter of Budday et al.~\cite{budday2017} serves as a benchmark for the hyperelastic model, this material indeed behaves highly viscoelastic, and the pseudo-elastic stress-deformation tuples as shown in Fig.~\ref{fig:cortex_fit} have actually been obtained as an average of the experimental loading and unloading curves. 
Therefore, for a reliable mechanical model of this biomaterial, it would actually be indispensable to take the viscous effects into account, see~\cite{ricker2023a}.}
Pseudo-elastic data from the UT, UC and SS experiments of~\cite{budday2017} and the corresponding models are shown in Fig.~\ref{fig:cortex_fit}.%
\footnote{For a tabular overview on the pseudo-elastic stress-deformation tuples obtained from the experiments, the reader is referred to~\cite[Table~1]{linka2023}.}
All the NN-based model variants are parameterised from UT and UC, whereas unseen SS data serves for the validation.
Overall, the performance of the PANN is almost independent of whether only $\ipci$ or $\ipcii$ or both $\ipci$ and $\ipcii$ are considered as model input.
In each case, good qualitative agreement is reached between fit and experimental UT/UC data, and this is also true for the SS prediction, apart from very large shear deformations, where a locking-like effect is predicted in several cases that is not in agreement with the experiments.
However, noticeable quantitative deviations are observed.
Such observations have also been made with alternative model ansatzes, see e.g.~\cite{linka2023,kuhl2024}. Two reasons for this seem plausible.
Firstly, there is a significant scatter in the experiments on biomaterials. Taking the uncertainty of the pseudo-elastic experimental data into account, the agreement with the model is very good.
Secondly, human cortex actually is a not truly elastic material. Instead, significant viscous effects are present even in the data under consideration. Therefore, it is doubtful whether a polyconvex hyperelastic model should be able to correctly capture the pseudo-elastic response.
Interestingly, an unconstrained FNN depending on both $\ipci$ and $\ipcii$ enables an excellent quantitative agreement for all three experiments UT, UC, SS.

Different from the benchmarks on rubber elasticity presented in Sect.~\ref{sec:disc_trel}~and~\ref{sec:disc_alex}, both one invariant-only models work almost as well as the approach based on both $\ipci$ and $\ipcii$.
This effect can be attributed to the small amounts of deformation that are present in the cortex data set --~different from the experiments on the elastomers.
As shown in Fig.~\ref{fig:admInv}, there are only small differences between the invariants $\ipci$ and $\ipcii$ in this case.

\subsection{Alexander's data on neoprene}
\label{sec:disc_alex}

\begin{figure}[tbp!]
	\centering
	\begin{subfigure}{\linewidth}
		\centering
		\includegraphics[width=0.9\linewidth,trim=0mm 0mm 0mm 10mm]{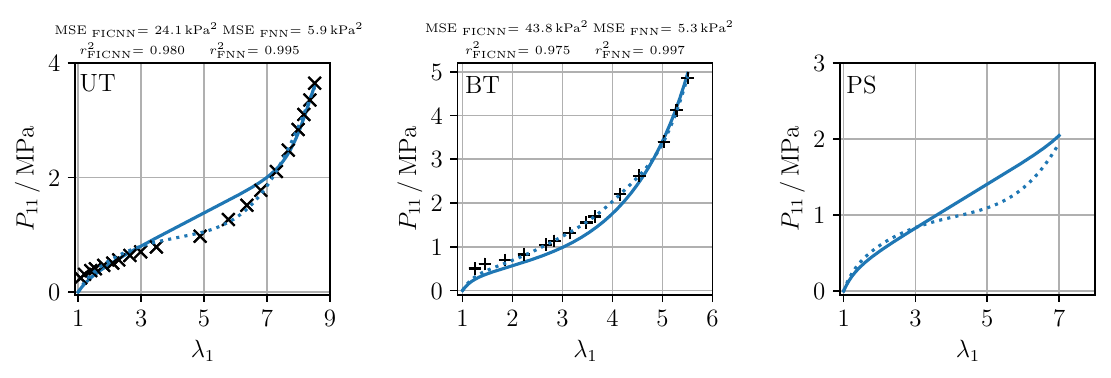} 
		\caption{Model including both $\ipci$ and $\ipcii$}		
		\label{fig:alex_both}
	\end{subfigure}
	\begin{subfigure}{\linewidth}
		\centering
		\includegraphics[width=0.9\linewidth,trim=0mm 0mm 0mm 0mm]{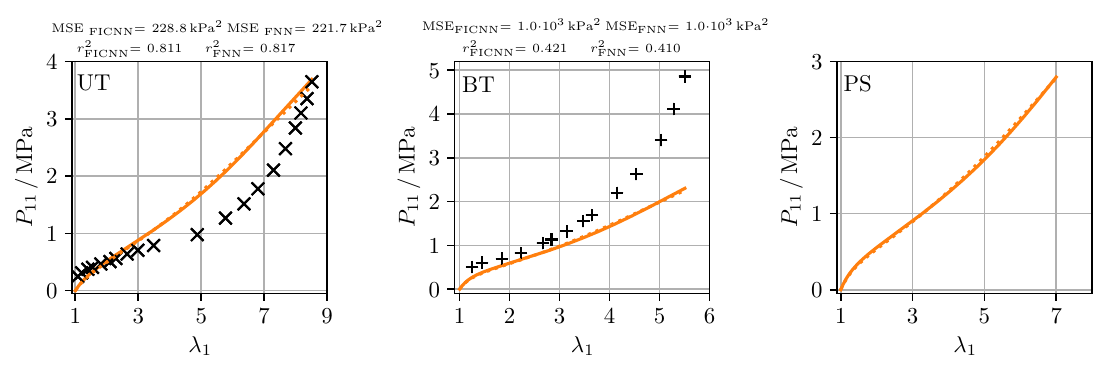} 
		\caption{Model including only $\ipci$. Note that the dotted lines overlap with the solid ones in a large stretch interval.}		
		\label{fig:alex_i1}
	\end{subfigure}
	\begin{subfigure}{\linewidth}
		\centering
		\includegraphics[width=0.9\linewidth,trim=0mm 0mm 0mm 0mm]{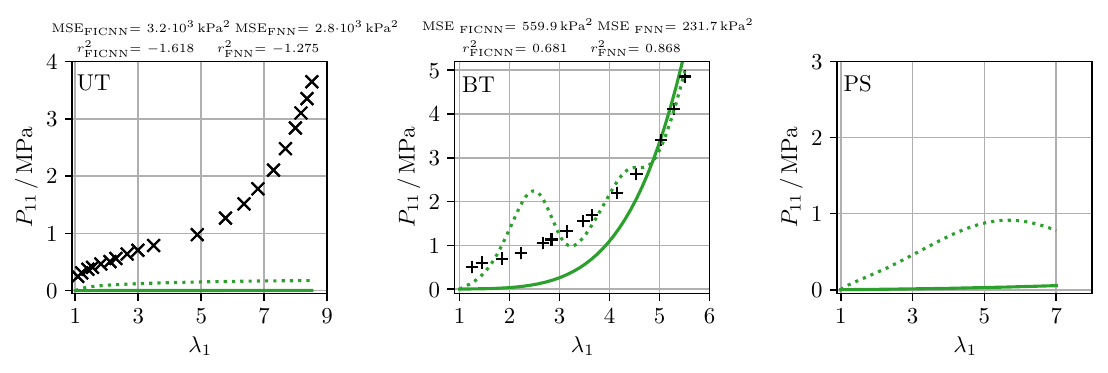} 
		\caption{Model including only $\ipcii$}		
		\label{fig:alex_i2}
	\end{subfigure}
	\includegraphics[scale=0.8,trim=0mm 0mm 5mm 0mm]{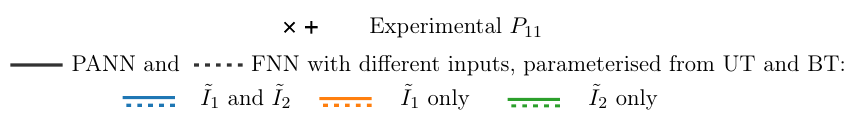}		
	\caption{Performance of the PANN (based on an FICNN) and an unconstrained FNN with respect to Alexander's experimental results for neoprene~\cite{alexander1968} depending on the invariants considered as input:
	either both $\ipci$ and $\ipcii$~(a) or only $\ipci$~(b) or solely $\ipcii$~(c) are considered.
	Results for combined fit of data for uniaxial tension~(UT) and equi-biaxial tension~(BT) ($N^\nv{UT}= N^\nv{BT} = 1$) and predictions for pure shear~(PS).
	}
	\label{fig:alex_fit}
\end{figure}

As a final benchmark, the experimental results on UT and BT of neoprene by Alexander~\cite[Fig.~4--5]{alexander1968} are investigated. For convenience, we provide these data in the Appendix in Tab.~\ref{table:alexanderData}.
The models are parameterised simultaneously from UT and BT ($N^\nv{UT}= N^\nv{BT} = 1$). Since experimental results for PS are not available, only model predictions are shown for this deformation scenario.
Overall, the results are similar to the findings regarding Treloar's natural rubber in Sect.~\ref{sec:disc_trel}, which confirms the conclusions on the role of $\ipci$ and $\ipcii$ drawn above. For the models considering both $\ipci$ and $\ipcii$, good quantitative agreement between PANN and experiment is reached for UT and BT, and plausible predictions are made for PS, Fig.~\ref{fig:alex_both}. 
For the FNN, even better agreement is obtained between fit and experimental UT/BT data. However, there is a wave-like artefact in the predictions for PS similar to the FNN prediction to Treloar's data, which is not expected in experiments.
If the dependency of the material on $\ipcii$ is dropped, Fig.~\ref{fig:alex_i1}, qualitatively sound fits and predictions are possible. However, significant deviations between fit and experimental UT/BT are then observed, which are much more pronounced than for the NR-S investigated by Treloar.
In contrast, the behaviour of the $\ipcii$-only model, Fig.~\ref{fig:alex_i2}, is similar to the benchmark of Treloar. If $\ipci$ is not considered in the model, neither the PANN nor an unconstrained FNN does reflect any of the experiments correctly.

\section{Conclusions and outlook}
\label{sec:concl}

In this work, we provide an explicit representation of the set of admissible deformation invariants in incompressible elasticity, which is a semi-infinite surface in $\rset^2$. Furthermore, we prove that uniaxial and equi-biaxial deformation states correspond to the boundary of the set of admissible invariants.
As a consequence, suitable models parameterised from uniaxial and equi-biaxial tension, for instance, may often make correct predictions for other load cases, which is demonstrated for PANNs using representative benchmark data sets.
Moreover, we have demonstrated that considering only one invariant --~either $\ipci$ or $\ipcii$~-- can allow for good agreement with experiments in case of small deformations. In contrast, it is necessary to consider both invariants for precise models at larger amounts of strain.
However, when only uniaxial experimental data is available for model calibration, $\ipci$-only approaches are to be preferred over models including both $\ipci$ and $\ipcii$ in order to enable unphysically stiff predictions for general deformation scenarios.
Different from formulations including $\ipcii$, models based solely on $\ipci$ can make sound predictions for multiaxial loadings even if parameterised only from uniaxial data. On the contrary, $\ipcii$-only models are completely incapable in even qualitatively capturing experimental stress data at large deformations.

Since multiaxial loadings are essential for the adequate parameterisation of precise models including $\ipci$ and $\ipcii$, it seems promising to spend future work on combining PANN material models with full-field measurement-based methods such as the \textit{virtual fields method}~\cite{pierron2012} or the \textit{EUCLID} framework~\cite{flaschel2021,thakolkaran2022} and the \textit{mCRE} approach~\cite{benady2024,nguyen2025}.
In particular, unsupervised learning of the models from experiments with heterogeneous specimens could be carried out, cf.~\cite{jailin2024}. For this, setups should be designed such that a preferably representative set of deformation states can be covered.

\section*{Supplementary material}
After publication of the final version of the manuscript, an implementation of all the models considered in this work and the specific parameters for the examples discussed in Sect.~\ref{sec:fits} as well as scripts for the visualisation of the set of the admissible invariants will be available at \url{https://github.com/NEFM-TUDresden}.
	
\section*{Acknowledgements}
Support for this work was provided by the German Research Foundation (DFG) under grant KA 3309/20-1 (project 517438497).


\section*{CRediT author contribution statement}
\textit{Franz Dammaß:} Conceptualisation, Formal analysis, Investigation, Methodology, Visualisation, Software, Validation, Writing –- original draft, Writing –- editing, Funding acquisition.
\textit{Karl A. Kalina:} Conceptualisation, Software, Writing –- editing.
\textit{Markus Kästner:} Conceptualisation, Writing –- editing, Resources, Funding acquisition.

\section*{Declaration of competing interest}
There is no conflict of interest to declare.

\appendix

\section{Additional mathematical relationships}

\subsection{Positivity of the eigenvalues}
\label{sec:posEV}

\begin{lemma}
	Let $\kappa_1$, $\kappa_2$, $\kappa_3$ real solutions of the characteristic equation~\eqref{eq:charpol}, i.e. real eigenvalues of $\ciso$ and $\biso$.
	Then all eigenvalues are positive: $\kappa_m >0 \,\, \forall \, m \in \{1,2,3\}$.
	\label{lem:posEV}
\end{lemma}
\begin{proof}
	The proof can be established similarly to~\cite[Theorem~C.3]{linden2023} as follows: 
	From 
	\begin{equation}
		1 = \iiici = \kappa_1 \, \kappa_2 \, \kappa_3 \comma
		\label{eq:Jpos}
	\end{equation}
	it follows that either all eigenvalues are positive or two eigenvalues are positive and the other two are negative.
	Let us consider the latter case and, without loss of generality, assume $\kappa_1>0$ and $\kappa_2, \kappa_3<0$.
	From Lemma~\ref{lem:InvEstim}, we have $\ici >0$, which can be rewritten as
	\begin{equation}
		\kappa_1 > - (\kappa_2+\kappa_3)
		\comma
		\label{eq:kappa_1_lower}
	\end{equation}
	and $\iici >0$, i.e.
		\begin{equation}
		\kappa_1 \, (\kappa_2+\kappa_3) + \kappa_2 \, \kappa_3 >0 \Longleftrightarrow \kappa_1 < \frac{- \kappa_2 \, \kappa_3}{\kappa_2+\kappa_3}
		\point
		\label{eq:kappa_1_upper}
	\end{equation}
	Combining \eqref{eq:kappa_1_lower} and \eqref{eq:kappa_1_upper} leads to the contradiction
	\begin{equation}
		0 > (\kappa_2+\kappa_3)^2 - \kappa_2 \, \kappa_3 >0 
		\comma
	\end{equation}
	and therefore $\kappa_1, \kappa_2, \kappa_3 >0$.
\end{proof}

\subsection{Explicit expression for $f_\nv{out}(\ici)$}
\label{sec:explSolDZero}

Apart from $f_\nv{low}(\ici)$ and $f_\nv{up}(\ici)$, the following function is obtained as a third solution of $D(\ici,\iici)=0$:

\begin{align*}
	f_\nv{out}\left(\ici\right) =&
	-\frac{1}{24}
	\,\sqrt [3]{\ici^{6}-540\,\ici^{3}-5832+24\,\sqrt {-3\,\ici^{9}+243
			\,\ici^{6}-6561\,\ici^{3}+59049}}
	\\
	&+{\frac {6\, \left( -{3}/{2} \, \ici -{{\ici^{4}}/{144}} \right)}
			{\sqrt [3]{\ici^{6}-540\,{\ici}^{3}-5832+24\,\sqrt {-3\,\ici^{9}+243\,\ici^{6}-6561\,\ici^{3}+59049}}}
	 } + \frac{1}{12}\,\ici^{2}\\
	&+\frac{i}{2} \sqrt {3}
	\Bigg( 
	\frac{1}{12} \, \sqrt [3]{\ici^{6}-540\,\ici^{3}-5832+24\,\sqrt {
			-3\,\ici^{9}+243\,\ici^{6}-6561\,\ici^{3}+59049}}\\
	&\quad +12\,{\frac {1}{\sqrt [3
			]{\ici^{6}-540\,\ici^{3}-5832+24\,\sqrt {-3\,\ici^{9}+243\,\ici^{6}-6561\,
					\ici^{3}+59049}}} \left[ -\frac{3}{2} \, \ici -{\frac {\ici^{4}}{144}} \right] }
	\Bigg)
	\numberthis \label{eq:fout}
	\point
\end{align*}

\subsection{Derivatives of the invariants and stress tensor for simple shear (SS) deformation}
\label{sec:appPK1SS}

For an invariant-based incompressible model with $\psi(\te F, \pp) = \psii\left(\ici(\fiso), \iici(\fiso) \right) + \pp (J-1)$,
the derivatives
\begin{align}
	\diffp{\ici}{\te F} &= - \frac{2}{3} \, \ici \, \te F^{-\top} + 2 \, J^{-2/3} \, \te F
	\comma \label{eq:derivI1}\\
	\diffp{\iici}{\te F } &=  \frac{2}{3} \, \te F^{-\top} \iici
	- 2 \, J^{2/3} \, \te F^{-\top} \cdot \te{F}^{-1} \cdot \te F^{-\top}
	\label{eq:derivI2}
\end{align}
are required when evaluating the definition of stress \eqref{eq:PK1-red}.
Then, for a simple shear deformation, where
\[
\left[F_{kL}\right] \equiv \left[\overline{F}_{kL}\right] = 
\begin{bmatrix}
	1 & \gamma & 0\\ 0 & 1 & 0 \\ 0 & 0 & 1
\end{bmatrix} 
\]
with $\gamma \in \rset$,
the coordinates of the first Piola-Kirchhoff stress are obtained as
\begin{equation}
	{\footnotesize
	\left[P_{kL}\right]
	=
	\begin{bmatrix}
		-p - \gamma^2 \left( \dfrac{2}{3} \, \ddiffp{\psii}{\ici} + \dfrac{4}{3} \, \ddiffp{\psii}{\iici} \right)
		 & 2\, \gamma \left( \ddiffp{\psii}{\ici} + \ddiffp{\psii}{\iici} \right)
		 & 0\\
%
		\gamma \left( \gamma^{2} \, \left[ \dfrac{2}{3} \, \ddiffp{\psii}{\ici} +\dfrac{4}{3} \, \ddiffp{\psii}{\iici} \right]
		+ p+ 2\,\ddiffp{\psii}{\ici} + 2\,\ddiffp{\psii}{\iici} \right) 
		 &
		  -p - \gamma^2 \left( \dfrac{2}{3} \, \ddiffp{\psii}{\ici} + \dfrac{4}{3} \, \ddiffp{\psii}{\iici} \right) &
		   0 \\ 
		0 & 0 & 
		-p -\dfrac{2}{3} \,  \gamma^2 \left( \ddiffp{\psii}{\ici} - \ddiffp{\psii}{\iici} \right)
	\end{bmatrix} 
}
\label{eq:PK1_SS}
\end{equation}
with $p = -\pp$ denoting the hydrostatic pressure.

\begin{remark}
	For SS deformations and the stress tensor as provided in \eqref{eq:PK1_SS} there can be a difference between the normal stresses $P_{11}=P_{22}$ and $P_{33}$, which is non-zero:
	\begin{equation}
		P_{11}-P_{33}=P_{22}-P_{33}
		= -2 \, \diffp{\psii}{\iici}
	\end{equation}
	Similarly, in terms of the Cauchy stress $\tg \upsigma$ obtained from $\te P$ by \eqref{eq:pushForw}, we have
	\begin{equation}
		\sigma_{11}-\sigma_{33} = 2 \, \diffp{\psii}{\ici}
		\quad \glmand \quad
		\sigma_{22}-\sigma_{33} = -2 \, \diffp{\psii}{\iici}
		\point
	\end{equation}
	In the literature, these the normal stresses differences under SS deformation are referred to as \textit{Poynting-like effects} \cite{kuhl2024,horgan2012}, giving reference to \cite{poynting1909}.
\end{remark}

\section{Alexander's stress-deformation data}
\label{sec:alexRawData}

In Tab.~\ref{table:alexanderData}, we provide Alexander's experimental results on neoprene. These have been extracted from \cite[Figs.~4--5]{alexander1968}.
Cauchy stress $\tg \upsigma$ has been expressed in SI units and converted into the first Piola-Kirchhoff stress $\te P$ using relation~\eqref{eq:pushForw}.

\begin{table}[h!]
	\centering
	\caption{Experimental data on UT and BT of neoprene as published in \cite[Figs.~4--5]{alexander1968}}
	\begin{small}
	\begin{tabular}{c c|c c}
		\hline
		\multicolumn{2}{c|}{\textbf{UT}} & \multicolumn{2}{c}{\textbf{BT}} \\
		$\lambda_1$ & $P_{11} / (\text{N}/\text{mm}^2)$ & $\lambda_1$ & $P_{11} / (\text{N}/\text{mm}^2)$ \\
		\hline
		1.078 & 0.247 & 1.248 & 0.504 \\
		1.202 & 0.311 & 1.447 & 0.607 \\
		1.400 & 0.374 & 1.850 & 0.698 \\
		1.547 & 0.407 & 2.226 & 0.823 \\
		1.811 & 0.466 & 2.663 & 1.050 \\
		2.099 & 0.506 & 2.837 & 1.133 \\
		2.289 & 0.561 & 3.141 & 1.318 \\
		2.645 & 0.639 & 3.476 & 1.555 \\
		2.988 & 0.704 & 3.643 & 1.691 \\
		3.486 & 0.788 & 4.151 & 2.196 \\
		4.874 & 0.976 & 4.535 & 2.626 \\
		5.771 & 1.266 & 5.029 & 3.402 \\
		6.358 & 1.516 & 5.279 & 4.116 \\
		6.800 & 1.777 & 5.503 & 4.857 \\
		7.295 & 2.105 & & \\
		7.676 & 2.477 & & \\
		7.989 & 2.838 & & \\
		8.168 & 3.101 & & \\
		8.360 & 3.352 & & \\
		8.514 & 3.646 & & \\
		\hline
	\end{tabular}
	\label{table:alexanderData}
\end{small}
\end{table}

\bibliographystyle{abbrv}  
{\footnotesize \bibliography{bib_invarIncomp.bib}}

\end{document}